\documentclass[a4paper,11pt]{article}
\pdfoutput=1 

\usepackage{jheppub} 
\makeatletter
\DeclareRobustCommand*{\bfseries}{%
  \not@math@alphabet\bfseries\mathbf
  \fontseries\bfdefault\selectfont
  \boldmath
}
\makeatother

\usepackage{calc}
\usepackage{rotating}
\usepackage[english]{babel}
\usepackage{graphicx}
\usepackage{subfig}
\usepackage{float}
\usepackage{amsmath}
\usepackage{amssymb}
\usepackage{amsthm}
\usepackage{latexsym}
\usepackage{dcolumn}

\newcommand{\newc}{\newcommand*}

\long\def\begincomment#1\endcomment{%
        \begingroup\sf\baselineskip12pt#1\endgroup}

\newc{\etal}{\textrm{et al.}} 
\newc{\eg}{\textrm{e.g.}} 
\newc{\ie}{\textrm{i.e.}}
\newc{\etc}{\textrm{etc}}
\newc\vs{\textrm{vs.}}
\newc{\cl}{\rm {C.L.}}

\newc{\ev}{\ensuremath{\,\mathrm{eV}}}
\newc{\kev}{\ensuremath{\,\mathrm{keV}}}
\newc{\mev}{\ensuremath{\,\mathrm{MeV}}}
\newc{\gev}{\ensuremath{\,\mathrm{GeV}}}
\newc{\tev}{\ensuremath{\,\mathrm{TeV}}}
\newc{\MeV}{\mev} 
\newc{\TeV}{\tev}
\newc{\invpb}{\ensuremath{/\text{pb}}}
\newc{\invfb}{\ensuremath{/\text{fb}}}
\newc\nb{\ensuremath{\,\mathrm{nb}}} \newc\pb{\ensuremath{\,\mathrm{pb}}} \newc\fb{\ensuremath{\,\mathrm{fb}}}
\newc\pc{\ensuremath{\,\mathrm{pc}}}
\newc\kpc{\ensuremath{\,\mathrm{kpc}}}
\newc\mpc{\ensuremath{\,\mathrm{Mpc}}}
\newc\ps{\ensuremath{\,\mathrm{ps}}} 
\newc\cmeter{\ensuremath{\,\mathrm{cm}}} 
\newc\meter{\ensuremath{\,\mathrm{m}}} 
\newc\kmeter{\ensuremath{\,\mathrm{km}}}
\newc\second{\ensuremath{\,\mathrm{s}}}
\newc\msecond{\ensuremath{\,\mathrm{ms}}}
\newc\nsecond{\ensuremath{\,\mathrm{ns}}}
\newc\psecond{\ensuremath{\,\mathrm{ps}}}

\newc{\chisqmin}{\ensuremath{\chi^2_{\mathrm{min}}}}
\newc{\Delchisq}{\ensuremath{\Delta\chi^2}}
\newc{\chisq}{\ensuremath{\chi^2}}
\newc{\like}{\ensuremath{\mathcal{L}}}

\newc\lsim{\ensuremath{\mathrel{\rlap{\lower4pt\hbox{\hskip1pt$\sim$}}\raise1pt\hbox{$<$}}}}
\newc\gsim{\ensuremath{\mathrel{\rlap{\lower4pt\hbox{\hskip1pt$\sim$}}\raise1pt\hbox{$>$}}}}
\newc{\VEV}[1]{\ensuremath{\langle #1 \rangle}}
\newc{\dl}{\ensuremath{\stackrel{\leftarrow}{D}}}
\newc{\dr}{\ensuremath{\stackrel{\rightarrow}{D}}}

\newc{\bcenter}{\begin{center}}    \newc{\ecenter}{\end{center}}
\newc{\bfl}{\begin{flushleft}}    \newc{\efl}{\end{flushleft}}
\newc{\bfr}{\begin{flushright}}    \newc{\efr}{\end{flushright}}

\newc{\bi}{\begin{itemize}}
\newc{\ei}{\end{itemize}}
\newc{\bed}{\begin{description}}
\newc{\eed}{\end{description}}
\newc{\ben}{\begin{enumerate}}
\newc{\een}{\end{enumerate}}

\newc{\be}{\begin{equation}}
\newc{\ee}{\end{equation}}
\newc{\bea}{\begin{eqnarray}}
\newc{\eea}{\end{eqnarray}}
\newc{\ra}{\rightarrow}

\newc{\alphas}{\ensuremath{\alpha_s}}
\newc{\alphatwo}{\ensuremath{\alpha_2}}
\newc{\alphaone}{\ensuremath{\alpha_1}}
\newc{\alphai}[1]{\ensuremath{\alpha_{#1}}}
\newc{\alphaem}{\ensuremath{\alpha_{\mathrm{em}}}}
\newc{\alphaeff}{\ensuremath{\alpha_{\mathrm{eff}}}}
\newc{\sineff}{\ensuremath{\sin \theta_{\mathrm{eff}}}}
\newc{\sinsqeff}{\ensuremath{\sin^2 \theta_{\mathrm{eff}}}}
\newc{\dalphahad}{\ensuremath{\Delta \alpha_{\mathrm{had}}}}
\newc{\yt}{\ensuremath{h_t}} \newc{\yb}{\ensuremath{h_b}} \newc{\ytau}{\ensuremath{h_{\tau}}}
\newc\mz{\ensuremath{M_Z}} 
\newc\mw{\ensuremath{m_W}}
\newc\mZ{\mz}        \newc\mW{\mw}
\newc\mhsm{\ensuremath{ m_{H_{\mathrm{SM}}}}}
\newc{\mtop}{\ensuremath{ m_t}}               \newc{\mtpole}{\ensuremath{ M_t}}
\newc{\mbottom}{\ensuremath{ m_b}} 
\newc{\mtau}{\ensuremath{ m_{\tau}}}
\newc{\mt}{\mtpole}
\newc{\mb}{\mbottom} 
\newc{\rgg}{\ensuremath{R_{h}(\gamma\gamma)}}
\newc{\rzz}{\ensuremath{R_{h}(ZZ)}}
\newc{\rtwogg}{\ensuremath{R_{h_2}(\gamma\gamma)}}
\newc{\rtwozz}{\ensuremath{R_{h_2}(ZZ)}}
\newc{\ronegg}{\ensuremath{R_{h_1}(\gamma\gamma)}}
\newc{\ronezz}{\ensuremath{R_{h_1}(ZZ)}}
\newc{\rsiggg}{\ensuremath{R_{h_\textrm{sig}}(\gamma\gamma)}}
\newc{\rsigzz}{\ensuremath{R_{h_\textrm{sig}}(ZZ)}}
\newc{\llbar}{\ensuremath{\ell\bar{\ell}}}
\newc{\tauptaum}{\ensuremath{ \tau^+\tau^-}}
\newc{\qqbar}{\ensuremath{ q\bar{q}}} \newc{\ppbar}{\ensuremath{ p\bar{p}}}
\newc{\bbbar}{\ensuremath{ b\bar{b}}} \newc{\ttbar}{\ensuremath{ t\bar{t}}}
\newc{\ffbar}{\ensuremath{ f\bar{f}}} \newc{\tautaubar}{\ensuremath{ \tau\bar{\tau}}}
\newc{\mchi}{\ensuremath{m_{\chi}}}
\newc{\squark}{\ensuremath{\tilde{q}}}
\newc{\slepton}{\ensuremath{\tilde{l}}}
\newc{\gluino}{\ensuremath{\tilde{g}}} 
\newc{\mgluino}{\ensuremath{{m_{\gluino}}}}
\newc{\tone}{\ensuremath{{\tilde{t}_1}}}

\newc{\sthw}{\ensuremath{ \sin\theta_W}}              \newc{\cthw}{\ensuremath{\cos\theta_W}}
\newc{\tanthw}{\ensuremath{ \tan\theta_W}}              \newc{\cotthw}{\ensuremath{\cot\theta_W}}
\newc{\ssqthw}{\ensuremath{\sin^2 \theta_W}}
\newc{\msbar}{\ensuremath{\overline{MS}}} \newc{\drbar}{\ensuremath{\overline{DR}}}
\newc{\mtmtsmmsbar}{\ensuremath{ m_t(m_t)^{\msbar}_{{\mathrm{SM}}}}}
\newc{\mtmtsmdrbar}{\ensuremath{ m_t(m_t)^{\drbar}_{{\mathrm{SM}}}}}
\newc{\mtmtmssmdrbar}{\ensuremath{ m_t(m_t)^{\drbar}_{{\mathrm{SUSY}}}}}
\newc{\mbmbmsbar}{\ensuremath{ m_b(m_b)^{\msbar} }}
\newc{\mbmbsmmsbar}{\ensuremath{ m_b(m_b)^{\msbar}_{{\mathrm{SM}}}}}
\newc{\mbmzsmmsbar}{\ensuremath{ m_b(\mz)^{\msbar}_{{\mathrm{SM}}}}}
\newc{\mbmzsmdrbar}{\ensuremath{ m_b(\mz)^{\drbar}_{{\mathrm{SM}}}}}
\newc{\mbmzmssmdrbar}{\ensuremath{ m_b(\mz)^{\drbar}_{{\mathrm{SUSY}}}}}
\newc{\mtaumzsmmsbar}{\ensuremath{ m_{\tau}(\mz)^{\msbar}_{{\mathrm{SM}}}}}
\newc{\mtaumzsmdrbar}{\ensuremath{ m_{\tau}(\mz)^{\drbar}_{{\mathrm{SM}}}}}
\newc{\mtaumzmssmdrbar}{\ensuremath{ m_{\tau}(\mz)^{\drbar}_{{\mathrm{SUSY}}}}}
\newc{\alphasmzms}{\ensuremath{\alpha_s(M_Z)^{\overline{MS}}}}
\newc{\alphaimzms}[1]{\ensuremath{\alpha_{#1}(M_Z)^{\overline{MS}}}}
\newc{\alphaemmz}{\ensuremath{\alpha_{\mathrm{em}}(M_Z)^{\overline{MS}}}}

\newc{\mzero}{\ensuremath{{m_0}}}
\newc{\mhalf}{\ensuremath{ m_{1/2}}}
\newc{\tanb}{\ensuremath{\tan\beta}}
\newc{\azero}{\ensuremath{ A_0}}
\newc{\bzero}{\ensuremath{ B_0}}
\newc{\signmu}{\ensuremath{\rm{sgn}\,\mu}}
\newc{\mueff}{\ensuremath{\mu_{\rm{eff}}}}
\newc{\lam}{\ensuremath{{\lambda}}}
\newc{\kap}{\ensuremath{{\kappa}}}
\newc{\alam}{\ensuremath{{A_{\lambda}}}}
\newc{\akap}{\ensuremath{{A_{\kappa}}}}
\newc{\hs}{\ensuremath{ H_s}}      
\newc{\mhs}{\ensuremath{ m_{H_s}}} 
\newc{\mgut}{\ensuremath{ M_{\rm GUT}}}
\newc{\mplanck}{\ensuremath{ M_{\rm P}}}      \newc{\mpl}{\ensuremath{ M_{\rm Pl}}}
\newc{\msusy}{\ensuremath{ M_{\rm SUSY}}}      \newc{\ms}{\ensuremath{ M_{\rm S}}}
 \newc{\mhl}{\ensuremath{m_\hl}} 
 \newc{\mhone}{\ensuremath{m_{h_1}}} 
 \newc{\mhtwo}{\ensuremath{m_{h_2}}} 
 \newc{\mglu}{\ensuremath{m_{\tilde g}}} 
 \newc{\mul}{\ensuremath{m_{\tilde{u}_L}}} 
 \newc{\mtone}{\ensuremath{m_{\tilde{t}_1}}} 
 \newc{\ma}{\ensuremath{m_A}} 
 \newc{\maone}{\ensuremath{m_{a_1}}} 
 \newc{\matwo}{\ensuremath{m_{a_2}}}
 \newc{\hone}{\ensuremath{h_1}}
 \newc{\htwo}{\ensuremath{h_2}}
 \newc{\aone}{\ensuremath{a_1}}
 \newc{\atwo}{\ensuremath{a_2}}
 \newc{\mhu}{\ensuremath{ m_{H_u}}}       
 \newc{\mhd}{\ensuremath{ m_{H_d}}}
 \newc{\mhusq}{\ensuremath{ m_{H_u}^2}}       
 \newc{\mhdsq}{\ensuremath{ m_{H_d}^2}}
 \newc{\mhuew}{\ensuremath{ m^{\ast}_{H_u}}}       
 \newc{\mhdew}{\ensuremath{ m^{\ast}_{H_d}}}
 \newc{\mhuewsq}{\ensuremath{ m^{\ast\, 2}_{H_u}}}       
 \newc{\mhdewsq}{\ensuremath{ m^{\ast\, 2}_{H_d}}}
 \newc{\hu}{\ensuremath{ H_u}}       
 \newc{\hd}{\ensuremath{ H_d}}

 \newc{\mqthree}{\ensuremath{m_{\widetilde{Q}_3}^2}}
 \newc{\muthree}{\ensuremath{m_{\tilde{u}_3}^2}}
 \newc{\mdthree}{\ensuremath{m_{\tilde{d}_3}^2}}
 \newc{\mlthree}{\ensuremath{m_{\widetilde{L}_3}^2}}
 \newc{\methree}{\ensuremath{m_{\tilde{e}_3}^2}}
 \newc{\mqtwo}{\ensuremath{m_{\widetilde{Q}_2}^2}}
 \newc{\mutwo}{\ensuremath{m_{\tilde{u}_2}^2}}
 \newc{\mdtwo}{\ensuremath{m_{\tilde{d}_2}^2}}
 \newc{\mltwo}{\ensuremath{m_{\widetilde{L}_2}^2}}
 \newc{\metwo}{\ensuremath{m_{\tilde{e}_2}^2}}
 \newc{\mqone}{\ensuremath{m_{\widetilde{Q}_1}^2}}
 \newc{\muone}{\ensuremath{m_{\tilde{u}_1}^2}}
 \newc{\mdone}{\ensuremath{m_{\tilde{d}_1}^2}}
 \newc{\mlone}{\ensuremath{m_{\widetilde{L}_1}^2}}
 \newc{\meone}{\ensuremath{m_{\tilde{e}_1}^2}}
 \newc{\mone}{\ensuremath{M_1}}
 \newc{\monesq}{\ensuremath{M_1^2}}
 \newc{\mtwo}{\ensuremath{M_2}}
 \newc{\mtwosq}{\ensuremath{M_2^2}}
 \newc{\mthree}{\ensuremath{M_3}}
 \newc{\mthreesq}{\ensuremath{M_3^2}}
 \newc{\atau}{\ensuremath{{A_{\tau}}}}
 \newc{\at}{\ensuremath{{A_{t}}}}
 \newc{\ab}{\ensuremath{{A_{b}}}}
 \newc{\atausq}{\ensuremath{{A_{\tau}^2}}}
 \newc{\atsq}{\ensuremath{{A_{t}^2}}}
 \newc{\absq}{\ensuremath{{A_{b}^2}}}

 \newc{\dmzero}{\ensuremath{\Delta{_{m_0}}}}
 \newc{\dmhalf}{\ensuremath{\Delta{_{m_{1/2}}}}}
 \newc{\dmu}{\ensuremath{\Delta{_{\mu}}}}

 \newc{\pten}{\ensuremath{\psi_{10}}}
 \newc{\ffive}{\ensuremath{\phi_{5}}}
 \newc{\hfive}{\ensuremath{h_{5}}}
 \newc{\hbfive}{\ensuremath{h_{\bar{5}}}}
 \newc{\thet}{\ensuremath{\theta_{50}}}
 \newc{\thetb}{\ensuremath{\theta_{\,\overline{50}}}}
 \newc{\ptenhat}{\ensuremath{\hat{\psi}_{10}}}
 \newc{\ffivehat}{\ensuremath{\hat{\phi}_{5}}}
 \newc{\hfivehat}{\ensuremath{\hat{h}_{5}}}
 \newc{\hbfivehat}{\ensuremath{\hat{h}_{\bar{5}}}}
 \newc{\thethat}{\ensuremath{\hat{\theta}_{50}}}
 \newc{\thetbhat}{\ensuremath{\hat{\theta}_{\,\overline{50}}}}
 \newc{\si}{\ensuremath{\Sigma}}
 \newc{\mfive}{\ensuremath{m_5^2}}
 \newc{\mten}{\ensuremath{m_{10}^2}}
 \newc{\dfive}{\ensuremath{\Delta^2_5}}
 \newc{\dbfive}{\ensuremath{\Delta^2_{\bar{5}}}}
 \newc{\dfifty}{\ensuremath{\Delta^2_{50}}}
 \newc{\dfiftyb}{\ensuremath{\Delta^2_{\,\overline{50}}}}
 \newc{\msi}{\ensuremath{m_{\Sigma}^2}}
 \newc{\lamh}{\ensuremath{\lambda_{H}}}
 \newc{\lamhb}{\ensuremath{\lambda_{\bar{H}}}}
 \newc{\ah}{\ensuremath{A_{H}}}
 \newc{\ahb}{\ensuremath{A_{\bar{H}}}}
 \newc{\lams}{\ensuremath{\lambda_{S}}}
 \newc{\as}{\ensuremath{A_{S}}}
 \newc{\lamsig}{\ensuremath{\lambda_{\si}}}
 \newc{\asig}{\ensuremath{A_{\si}}}

 \newc{\msten}{\ensuremath{m_{16}^2}}
 \newc{\mhun}{\ensuremath{m_{126}^2}}
 \newc{\mhunb}{\ensuremath{m_{\bar{126}}^2}}
 \newc{\mthun}{\ensuremath{m_{210}^2}}
 \newc{\ahun}{\ensuremath{A_{\bar{126}}}}
 \newc{\yhun}{\ensuremath{Y_{\bar{126}}}}
 \newc{\aten}{\ensuremath{A_{10}}}
 \newc{\yten}{\ensuremath{Y_{10}}}
 \newc{\alone}{\ensuremath{A_{\lambda_1}}}
 \newc{\altwo}{\ensuremath{A_{\lambda_2}}}
 \newc{\althree}{\ensuremath{A_{\lambda_3}}}
 \newc{\althreeb}{\ensuremath{A_{\bar{\lambda_3}}}}
 \newc{\lone}{\ensuremath{\lambda_1}}
 \newc{\ltwo}{\ensuremath{\lambda_2}}
 \newc{\lthree}{\ensuremath{\lambda_3}}
 \newc{\lthreeb}{\ensuremath{\bar{\lambda_3}}}

\newc{\sigsip}{\ensuremath{\sigma^{\rm SI}_{p}}}	\newc{\sigsin}{\ensuremath{\sigma^{\rm SI}_{n}}}
\newc{\sigsdp}{\ensuremath{\sigma^{\rm SD}_{p}}}	\newc{\sigsdn}{\ensuremath{\sigma^{\rm SD}_{n}}}
\newc{\sigsi}{\ensuremath{\sigma^{\rm SI}}}	\newc{\sigsd}{\ensuremath{\sigma^{\rm SD}}}
\newc{\abund}{\ensuremath{ \Omega h^2}}
\newc{\omegadm}{\ensuremath{ \Omega_{{\rm DM}}}}     \newc{\abunddm}{\ensuremath{ \Omega_{{\rm DM}} h^2}} 
\newc{\omegam}{\ensuremath{ \Omega_{{\rm m}}}}       \newc{\abundm}{\ensuremath{ \Omega_{{\rm m}} h^2}}
\newc{\omegab}{\ensuremath{ \Omega_{{\rm b}}}}	\newc{\abundb}{\ensuremath{ \Omega_{{\rm b}} h^2}}
\newc{\omegatot}{\ensuremath{ \Omega_{{\rm TOT}}}}
\newc{\omegacdm}{\ensuremath{ \Omega_{{\rm CDM}}}}   \newc{\abundcdm}{\ensuremath{ \Omega_{{\rm CDM}} h^2}}
\newc{\omegalambda}{\ensuremath{ \Omega_{\Lambda}}} \newc{\abundlambda}{\ensuremath{ \Omega_{\Lambda} h^2}}
\newc{\omegarad}{\ensuremath{ \Omega_{{\rm rad}}}}  \newc{\abundrad}{\ensuremath{ \Omega_{{\rm rad}} h^2}}
\newc{\rhocrit}{\ensuremath{ \rho_{\rm crit}}}
\newc{\rhochi}{\ensuremath{ \rho_{\chi}}}
\newc{\abunchi}{\ensuremath{\Omega_\chi h^2}}
\newc{\abundlsp}{\ensuremath{\Omega_{\rm LSP}h^2}}
\newcommand*{\abundchi}{\ensuremath{\Omega_\chi h^2}}

\newc{\amu}{\ensuremath{ a_{\mu}}}        \newc{\amususy}{\ensuremath{ a_{\mu}^{\mathrm{SUSY}}}}
\newc{\amuexpt}{\ensuremath{ a_{\mu}^{\mathrm{expt}}}}        \newc{\amusm}{\ensuremath{ a_{\mu}^{\mathrm{SM}}}}
\newc\deltaamu{\ensuremath{\Delta a_{\mu}}} \newc{\deltaamususy}{\ensuremath{\delta a_{\mu}^{\mathrm{SUSY}}}}
\newc\gmtwo{\ensuremath{ (g-2)_{\mu}}} 
\newc{\deltagmtwomususy}{\ensuremath{\delta\left(g-2\right)_{\mu}^{\mathrm{SUSY}}}}
\newc{\deltagmtwomu}{\ensuremath{\delta\left(g-2\right)_{\mu}}}
\newc\BR{\ensuremath{\rm BR}}
\newc\bsgamma{\ensuremath{ b\rightarrow s \gamma }}
\newc\bxsgamma{\ensuremath{\overline{B}\rightarrow X_{s}\gamma}}
\newc\brbsgamma{\ensuremath{\BR\left(\bsgamma\right)}}
\newc\brbxsgamma{\ensuremath{\BR\left(\bxsgamma\right)}}
\newc\bsmumu{\ensuremath{B_s\to\mu^+\mu^-}}
\newc\brbsmumu{\ensuremath{\BR\left(B_s\to\mu^+\mu^-\right)}}
\newc\bdmmumu{\ensuremath{\overline{B}_d\to\mu^+\mu^-}}
\newc\bbbarmix{\ensuremath{\overline{B}_s\mbox{-}B_s}}      
\newc\delmbs{\ensuremath{\Delta M_{B_s}}}
\newc{\butaunu}{\ensuremath{B_u \rightarrow \tau \nu}}
\newc{\brbutaunu}{\ensuremath{\BR\left(B_u \rightarrow \tau \nu\right)}}

\newcommand*{\reffig}[1]{Fig.~\ref{#1}}

     \newcommand*{\refsec}[1]{Sec.~\ref{#1}}

\newcommand*{\mstopone}{\ensuremath{m_{\tilde{t}_1}}}
\newcommand*{\mstoptwo}{\ensuremath{m_{\tilde{t}_2}}}

\newcommand*{\softsusy}{SOFTSUSY}
\newcommand*{\feynhiggs}{FeynHiggs}
\newcommand*{\higgsbounds}{H{\scriptsize IGGS}B{\scriptsize OUNDS}}
\newcommand*{\higgssignals}{H{\scriptsize IGGS}S{\scriptsize IGNALS}}
\newcommand*{\micromegas}{MicrOMEGAs}
\newcommand*{\multinest}{MultiNest}

\newcommand*{\superiso}{\text{SuperIso}}

\let\oldcite\cite
\renewcommand*{\cite}{~\oldcite}

\newcommand*{\hl}{\ensuremath{h}}

\restylefloat{figure}

\title{Low fine tuning in the MSSM with higgsino dark matter and unification constraints}

\author{Kamila Kowalska,} 
\author[1]{Leszek Roszkowski,\note{On leave of absence from the University of Sheffield, U.K.}}
\author{Enrico Maria Sessolo}
\author{and Sebastian Trojanowski}

\affiliation{National Centre for Nuclear Research,
  Ho{\. z}a 69, 00-681 Warsaw, Poland} 

\emailAdd{Kamila.Kowalska@fuw.edu.pl}
\emailAdd{L.Roszkowski@sheffield.ac.uk}
\emailAdd{Enrico-Maria.Sessolo@fuw.edu.pl}
\emailAdd{Sebastian.Trojanowski@fuw.edu.pl}

\abstract{We examine the issue of fine tuning in the MSSM with GUT-scale boundary conditions.
We identify specific unification patterns and mass relations that can lead to a significant lowering of the fine tuning due to gauginos, 
scalars, and the $\mu$ parameter, relative to the simplest unification conditions.
We focus on a phenomenologically interesting region that is favored by the Higgs mass and the relic density where the dark matter 
is a nearly pure higgsino with mass given by $\mu\simeq1\tev$ while the scalars and gauginos have masses in the multi-TeV regime. 
There, we find that the fine tuning can be reduced to the level of a few percent.
Despite the gluino mass in the ballpark of 2~TeV, resulting mass spectra will be hard to explore at the LHC, 
but good prospects for detection come from dark matter direct detection experiments. 
Finally, we demonstrate with a specific example how the conditions and mass relations giving low fine tuning 
can originate in the context of supergravity and Grand Unified Theories.}

\begin{document}
\maketitle
\flushbottom

\section{Introduction}\label{intro:sec}

The recent discovery of the Higgs boson with mass $\mhl\simeq
126\gev$\cite{Chatrchyan:2012ufa,Aad:2012tfa} was universally
acclaimed as a great experimental and theoretical success, as it
completed the verification of the Standard Model (SM) of particle
physics.  The mass of the new boson is relatively close to the scale
of electroweak (EW) interactions, $M_Z\simeq 91.2\gev$, consistent
with the predictions of low-scale supersymmetry (SUSY).

On the other hand, in the Minimal Supersymmetric Standard Model (MSSM)
the tree level value of the Higgs mass is bounded from above by $M_Z$,
and large radiative corrections are necessary to push the mass up to
the measured value.  This requires either a large SUSY scale,
$\msusy=(\mstopone\mstoptwo)^{1/2}\sim\textrm{several TeV}$, or
nearly maximal stop mixing, $|X_t|/\msusy\simeq\sqrt{6}$.  While the
above conditions can be easily satisfied in the MSSM and other
phenomenological models due to the freedom that comes with a large
number of parameters, in SUSY scenarios constrained by boundary
conditions at the scale of Grand Unification, \mgut, the available
parameter space is significantly reduced.  Particularly, in the
prototypical Constrained MSSM (CMSSM) it was shown that, by employing
two-loop calculations of the Higgs mass, the regions consistent with
the experimental value are most naturally characterized by \msusy\ in
the multi-TeV regime\cite{Akula:2012kk,Cabrera:2012vu,Kowalska:2013hha}.  Only
recently it was shown that including three-loop corrections pushes the available
parameter space to regions of somewhat lower, albeit still TeV-scale,
scalar masses\cite{Feng:2013tvd,Buchmueller:2013rsa}.

Since the imminent discovery of the Higgs boson was
anticipated, it was pointed out
in\cite{Papucci:2011wy,Hall:2011aa,Baer:2012uy,CahillRowley:2012rv}
and many other successive papers that the ensuing large values of \msusy\ put
SUSY at odds with requirements of EW naturalness (for a
non-comprehensive list of early articles on this vast topic
see\cite{Ellis:1986yg,Barbieri:1987fn,Ross:1992tz,Carlos:1993yy,Anderson:1994dz,Anderson:1994tr,Dimopoulos:1995mi,Chankowski:1997zh,
Giusti:1998gz,Barbieri:1998uv,Wright:1998mk,Chankowski:1998xv,Chacko:2005ra,Choi:2005hd,Chan:1997bi,Feng:1999mn,Feng:1999zg,
Kane:1998im,BasteroGil:1999gu,Kane:2002ap,Abe:2007kf,Horton:2009ed,Kitano:2005wc,Kitano:2006gv}).
In short, when SUSY is embedded into a more fundamental theory defined
at a very high scale, a large \msusy\ implies that the input parameters
have to be fine-tuned to uncomfortably high precision to
obtain the correct value of the EW symmetry breaking (EWSB) scale,
$v=246\gev$, when the parameters are run down to \msusy\ through the
Renormalization Group Equations (RGEs). However, scenarios of
\textit{a priori} natural SUSY, characterized by the stops with masses~$\lesssim600\gev$ and relatively low values of
the Higgs/higgsino mass parameter $\mu$\cite{Cohen:1996vb,Lodone:2012kp,Lee:2012sy,Cao:2012rz,Espinosa:2012in,Baer:2012uy},
become increasingly less viable as the limits on third generation
squarks from the LHC became more
stringent\cite{CahillRowley:2012kx,Chakraborty:2013moa,Boehm:2013qva,Buchmueller:2013exa,Han:2013kga,Kribs:2013lua,Kowalska:2013ica}.
In addition, the unavoidable requirement of maximal stop mixing, necessary
to fit the measured Higgs mass when the stop mass is not too large,
typically also leads to high fine
tuning\cite{Hall:2011aa,Kribs:2013lua,Kowalska:2013ica,Blanke:2013uia}.
   
To ameliorate the fine tuning problem, popular approches
involved, for example, adding extra sectors to the MSSM
in order to raise the value of the tree-level Higgs mass, thus allowing for 
lighter stop masses (like in the case of the Next-to-Minimal Supersymmetric 
SM (NMSSM)\cite{BasteroGil:2000bw,Delgado:2010uj,Ellwanger:2011mu,Ross:2011xv,Ross:2012nr} or other interesting models\cite{Athron:2013ipa}). 
Alternatively, in the context of models constrained at \mgut, one started to look 
for regions of the parameter space where ``cancellations" among the RGE running of the parameters
kept the fine tuning with respect to the high scale relatively low\cite{Chan:1997bi,Feng:1999mn,Feng:1999zg,Kane:1998im,BasteroGil:1999gu,Kane:2002ap,Abe:2007kf,Horton:2009ed,Antusch:2011xz,Antusch:2012gv,Baer:2012cf,
Baer:2012up,Baer:2012mv,Baer:2013xua,Baer:2013bba,Baer:2013gva}.    

The latter strategy resonates with the idea that the \textit{a priori} fine tuning does not necessarily entails an unnatural low-energy theory, 
but may rather be a result of our ignorance about the physics of the high scale\cite{Cassel:2009ps}, generically called $M_X$ hereafter. 
As a matter of fact, one may turn the argument around and use a ``bottom-up" approach to identify mass relations at the high scale 
that give low fine tuning in regions of low-scale SUSY that are favored by phenomenology and that would otherwise 
be ``generically" characterized by high levels of fine tuning.

It is well known that different high-scale models can lead to different levels of fine tuning.
When quantified through the Barbieri-Giudice measure\cite{Ellis:1986yg,Barbieri:1987fn}, 
$\Delta=|\partial\ln M_Z^2/\partial\ln p^2|$, with $p$ denoting high-scale input parameters, 
fine tuning indicates excessive sensitivity of the EW scale to variations in the soft SUSY-breaking parameters defined at $M_X$. 
In  general, $\Delta$ increases with $\ln(M_X)$ and may become very large for models defined at the unification or Planck scale. 
However, if certain relations among the high-scale parameters exist due to 
some new physics at $M_X$, $\Delta$ can be significantly smaller than in the uncorrelated case,
as those relations propagate through RGE running and conspire to make the EW scale much less sensitive to $M_X$ inputs. 
This is the known mechanism of ``focusing", which has been widely 
studied in the literature, for scalars (focus-point (FP) SUSY\cite{Chan:1997bi,Feng:1999mn,Feng:1999zg,Feng:2011aa,Feng:2012jfa}) 
and gauginos alike (non-universal gaugino mass models (NUGM)\cite{Kane:1998im,BasteroGil:1999gu,Kane:2002ap,Abe:2007kf,Horton:2009ed,Antusch:2012gv,Gogoladze:2012yf,Gogoladze:2013wva,Kaminska:2013mya,Kaminska:2014wia}).
There have also been several attempts to relate the scalar and gaugino sectors, 
for example in the framework of higher-dimensional brane scenarios\cite{Mayes:2013bda,Brummer:2013dya}. 

A separate issue is the fine tuning due to the parameter $\mu$, which is in most models
a SUSY-conserving parameter and appears in the superpotential 
independently of the soft SUSY-breaking terms. In order to keep the fine tuning due to $\mu$ low 
and at the same time satisfy the condition of EWSB, one generally assumes $\mu\lesssim 200-300\gev$. 
A characteristic feature of ``low $\mu$" scenarios is, 
however, an under-abundance of neutralino dark matter with respect to the 
value measured by PLANCK\cite{Ade:2013zuv}. While this is not fatal for the model (one can conceive additional particles that
can boost the value of the relic density, as explained, e.g., in\cite{Choi:2013lwa} 
and references therein), one might feel reluctant to abandon the simplicity of having just one dark matter candidate.

So the question arises as to whether there exist ways of reconciling neutralino
dark matter with naturalness in SUSY models defined at $M_X\sim\mgut$.
As was mentioned above, recent global analyses of the CMSSM\cite{Akula:2012kk,Cabrera:2012vu,Kowalska:2013hha} 
and the Non-Universal Higgs Model\cite{Roszkowski:2009sm}
have shown that the Higgs mass and flavor constraints 
are well satisfied in an extended region of the parameter space with large scalar and gaugino masses,
over which the relic density constraint can also be satisfied thanks to a nearly pure higgsino 
lightest SUSY particle (LSP) with mass 
$\mchi\approx \mu\approx 1\tev$. This region was shown to appear in the phenomenological MSSM\cite{Profumo:2004at,Fowlie:2013oua} and the NMSSM\cite{Kaminska:2013mya,Kaminska:2014wia} as well. It provides an attractive solution in the direction of 
reconciling EW naturalness with neutralino dark matter:
it is extended, phenomenologically viable (except for \deltagmtwomu), it naturally gives
$\mhl\simeq 126\gev$ and $\Omega_{\chi}h^2\simeq 0.12$ and, when different focusing mechanisms are assumed 
to reduce the fine tuning of scalars and gauginos, it is very stable from the fine-tuning point of view.
In fact, $\mu\simeq 1\tev$ means a value of $\Delta\simeq 250$, which is quite large but not catastrophic. 

The goal of this paper is to investigate the extent to which relations among the 
MSSM parameters defined at \mgut\ can be used to reduce the 
level of fine tuning in the $\sim1\tev$ higgsino region. 
We will show that, if one includes the $\mu$ parameter on the same footing as the soft-breaking terms,
the specific mass relation that ensues induces a total fine tuning
$\Delta\simeq 20$, even when the scalar and gaugino masses 
are in the multi-TeV regime and $\mu\simeq1\tev$. 
We will show that this relation emerges naturally in high-scale physics scenarios well known in the literature,
like the supergravity mechanism first introduced by Giudice and Masiero\cite{Giudice:1988yz} 
combined with the Missing Partner mechanism\cite{Masiero:1982fe,Grinstein:1982um} in Grand Unified Theories (GUTs).
While such scenarios have their problems, our goal is to provide an example of how the relations we determine
in the bottom-up approach might be used by the model building community in 
their effort to solve the puzzle of MSSM naturalness.

This paper is organized as follows. In \refsec{sec:ft} we define the measure we use to quantify the fine tuning
and describe the numerical approach to calculate its value.
In \refsec{sec:gut_fine} we derive the GUT-scale relations that can reduce the fine
tuning in the gaugino and scalar sectors of some specific, constrained models.  
In \refsec{sec:muterm} we show that there exist experimentally favored regions in the parameter space 
of these models for which relating $\mu$ and the scalar sector can lead to
further substantial reduction of the fine tuning. We show that this can be achieved through the Giudice-Masiero mechanism.
We give examples of possible physical spectra and phenomenology.  
In \refsec{sec:model} we show that the Missing Partner mechanism can be employed to complete the picture that  
leads to the desired fine-tuning properties.
Finally we present our summary and conclusions in \refsec{sec:con}.

\section{Definition of fine tuning}\label{sec:ft}

The criterion of fine tuning is intrinsically subjective and difficult to quantify in a unique way.
Different definitions and measures exist in the literature\cite{Ellis:1986yg,Barbieri:1987fn,Anderson:1994dz,Anderson:1994tr,Kitano:2005wc,Kitano:2006gv,Baer:2013bba,Baer:2013gva,Cabrera:2008tj,Fichet:2012sn,Farina:2013mla}
and the amount of fine tuning for a particular model can differ
when different measures are employed.
Throughout this paper we will use the well-known Barbieri-Giudice measure\cite{Ellis:1986yg,Barbieri:1987fn}:
$\Delta=\max \{\Delta_{p_i}\}$, where
\be 
\Delta_{p_i}=\left|\frac{\partial\ln M_Z^2}{\partial\ln p_i^2}\right|=\frac{1}{2}\left|\frac{\partial\ln M_Z^2}{\partial\ln p_i}\right|\,,
\label{finetune}
\ee 
and $p_i$ are the defining parameters of the model.
We do not assign absolute meaning to the numerical value of $\Delta$,
but rather take it as an estimate of the relative fine tuning of considered models.

$\Delta$ quantifies the stability of the global minimum of the scalar potential, 
$v^2\equiv M_Z^2/g^2$ (where $g^2\equiv (g_1^2+g_2^2)/2$ is the average of the $U(1)_Y$ and $SU(2)_L$ squared gauge couplings),
with respect to variations of the input parameters $p_i$.
Assuming that softly-broken SUSY is the low-scale remnant of a high-scale theory, like some GUT theory, or supergravity, or some string theory, 
the $p_i$'s are the parameters of an effective theory defined at the scale of gauge coupling unification, \mgut,
and they are renormalized through the RGEs to \msusy.
If the input parameters are independent from one another, $\Delta_{p_i}$ must be calculated for each of them 
separately and $\Delta$ often becomes significant.

The scale $M_Z$ is related to the other parameters through the well-known EWSB conditions that come from minimization
of the scalar potential. 
In this paper we will limit ourselves to the regions of $\tanb\geq 10$, where \tanb\ is the ratio of the Higgs doublets' 
vacuum expectation values (vev), as it will be clear below that regions of 
large \tanb\ can more easily show lower levels of fine tuning. 

In fact, for large \tanb\ the EWSB conditions read
\begin{eqnarray} 
\frac{M_Z^2}{2}&\approx &-\mu^2-\mhu^2-\Sigma_u^u+\mathcal{O}(\mhdsq/\tan^2\beta)\label{EWSBsim1}\,,\\
\frac{1}{\tanb}&\approx&\frac{B\mu-\Sigma_u^d}{\mhusq+\Sigma_u^u+\mhdsq+\Sigma_d^d+2\mu^2}\,,
\label{EWSBsim2}
\end{eqnarray}
where \mhu\ and \mhd\ are the soft-breaking masses of the Higgs doublets, 
$B\mu$ is the soft-breaking bilinear parameter, and  
the $\Sigma$ terms on the r.h.s. of Eqs.~(\ref{EWSBsim1}) and (\ref{EWSBsim2})
are the one-loop tadpole corrections to the scalar potential
whose expression in terms of physical masses and low-scale
soft-breaking parameters is given, e.g., in the Appendix of\cite{Baer:2012cf}.
The r.h.s.~of Eq.~(\ref{EWSBsim1})
is approximately independent of \tanb\ and, consequently, of the parameter $B\mu$.

In a theory defined in terms of a certain number of high-scale masses and trilinear couplings, called $p_i$,
that are subsequently run down to \msusy, one can integrate the RGEs to express
the low-scale parameters in terms of the high-scale ones. 
These expressions take the form of polynomial expansions,
e.g., $m_i(\msusy)=\sum_{ij} C_{ij} p_i p_j$, where the coefficients $C_{ij}$ depend on the running of the parameters 
$p_i$ between the two scales\cite{Martin:1993zk}.
As is common practice (see, e.g.,\cite{Antusch:2012gv}), we will explicitly write down the dependence
of $\mhusq$ at \msusy\ in terms of the GUT-scale masses and trilinear couplings:
\begin{flalign}\label{poli}
\mhusq(\msusy)&=0.645\mhusq+0.028\mhdsq-0.024\mqone-0.024\mqtwo-0.328\mqthree\nonumber\\
&+0.049\muone+0.049\mutwo-0.251\muthree-0.024\mdone-0.024\mdtwo-0.019\mdthree\nonumber\\
&+0.024\mlone+0.024\mltwo+0.024\mlthree-0.025\meone-0.025\metwo-0.025\methree\nonumber\\
&+0.014\monesq+0.210\mtwosq-1.097\mthreesq+0.001\mone\mtwo-0.047\mone\mthree-0.089\mtwo\mthree\nonumber\\
&-0.113\atsq+0.010\absq+0.006\atausq+0.008\at\ab+0.005\at\atau+0.004\ab\atau\nonumber\\
&+\mone(0.007\at-0.005\ab-0.004\atau)+\mtwo(0.062\at-0.009\ab+0.005\atau)\nonumber\\
&+\mthree(0.295\at+0.024\ab+0.030\atau).
\end{flalign}
The soft scalar masses and trilinear terms on the r.h.s.~of Eq.~(\ref{poli}) are all evaluated at the GUT scale.
To obtain the values of the coefficients in Eq.~(\ref{poli}) we used \texttt{\softsusy\ v3.3.9}\cite{softsusy} to expand around the CMSSM
point $\mzero=2\tev$, $\mhalf=1\tev$, $\azero=-1\tev$, $\tanb=30$. Since the coefficients of the expansion depend 
only on the RGE running of the masses, they change over the 
MSSM parameter space only by 10--20\%. Note, however, that such 
fluctuations can induce relatively large inaccuracies when Eq.~(\ref{poli}) is used indiscriminately to calculate the fine tuning 
over a large parameter space. 
Our results and conclusions below will be based on numerical calculations and we will use 
expansions like Eq.~(\ref{poli}) for semi-analytic considerations only.

The stability of $M_Z$ with respect to variations of the parameters $p_i$
is calculated by taking the derivative of Eq.~(\ref{EWSBsim1}) in the way of Eq.~(\ref{finetune}),
\begin{eqnarray}
\frac{\partial\ln M_Z^2}{\partial\ln p_i^2}&\approx & \frac{\partial}{\partial\ln p_i^2}\ln[-\mu^2-\mhusq(\msusy)-\Sigma_u^u(\msusy)]\nonumber\\
 &=&2\frac{p_i^2}{M_Z^2}\left[-\frac{\partial\mu^2}{\partial p_i^2}-\frac{\partial\mhusq(\msusy)}{\partial p_i^2}-\frac{\partial\Sigma_u^u(\msusy)}{\partial p_i^2}\right]\,.\label{natur}
\end{eqnarray}

The parameter $\mu$, commonly defined at \msusy, is related to its GUT-scale value, $\mu_0$, 
through RGE running, which depends only on the Yukawa couplings of the third generation, $y_{t,b,\tau}$,
and the gauge couplings of the $SU(2)_L$ and $U(1)_Y$ groups, $g_2$ and $g_1$: 
\be 
\mu = R(y_t,y_b,y_{\tau},g_1,g_2,\mgut/\msusy)\mu_0\approx 0.9\mu_0\,.\label{muscale}
\ee
On the other hand, $\partial\mhusq(\msusy)/\partial\mu_0^2=\partial \Sigma_u^u(\msusy)/\partial\mu_0^2=0$, 
so that $\Delta_{\mu}$ is given approximately by
\begin{equation}
\frac{\partial\ln M_Z^2}{\partial\ln \mu_0^2}\approx -2\frac{R^2\mu_0^2}{M_Z^2}=\frac{-2\mu^2}{M_Z^2}\,.\label{naturb}
\end{equation}
This shows the  well-known fact that in the MSSM naturalness requires preferably small values of $\mu = R\mu_0$\,.

The impact of any trilinear term or mass other than $\mu$ 
depends on the expansion of $\mhusq(\msusy)$, given approximately by Eq.~(\ref{poli}), or on the tadpole corrections.
This can be seen from Eq.~(\ref{natur}) by setting $\partial\mu^2/\partial p_i^2=0$.
The $\Sigma_u^u$ and other tadpole corrections depend indirectly on the high-scale parameters.
Applying the expressions of\cite{Baer:2012cf} to the case of small stop mixing, $|A_t|/\msusy\ll 1$, 
to which we limit ourselves in this study to minimize the fine tuning due to trilinear terms, $\Sigma_u^u$ reads 
\begin{eqnarray}
\Sigma_u^u(\mstopone,\mstoptwo)&\approx &\frac{3}{16 \pi^2}y_t^2\left[\mstopone^2\left(\frac{\mstopone^2-\mstoptwo^2}{\mstopone^2+\mstoptwo^2}\right)+\mstoptwo^2\left(\frac{\mstoptwo^2-\mstopone^2}{\mstopone^2+\mstoptwo^2}\right)-\mqthree-\muthree\right]\nonumber\\
 &\approx &\frac{3}{16 \pi^2}y_t^2\left(\frac{|\mqthree-\muthree|^2}{\mqthree+\muthree}
 -\mqthree-\muthree\right)\approx -\frac{3}{16 \pi^2}y_t^2\left(\mqthree+\muthree\right)\,,\label{sigmau}
\end{eqnarray}
where we used $\mstopone^2+\mstoptwo^2\approx \mqthree+\muthree\approx 2\msusy^2$, and 
$\mstoptwo^2-\mstopone^2\approx |\mqthree-\muthree|$, valid for relatively small mixing,
and all quantities are given at \msusy.
One can expand the low-energy
quantities in terms of the high-energy parameters, similarly to Eq.~(\ref{poli}), and quantify the fine tuning. 
We do not show this expansion here, since the contribution of tadpole terms to the fine-tuning level is subdominant, 
even in the regions of parameter space characterized 
by large scalar masses: $\partial \Sigma_u^u(\msusy)/\partial p_i^2\simeq 0.01\,\partial\mhusq(\msusy)/\partial p_i^2$
for broad ranges of $p_i$ values.
Nevertheless, in our numerical calculation of $\Delta$ we include also the contributions of the tadpoles, for which we 
have modified \texttt{\softsusy}.

Numerically, once a procedure for calculating the fine tuning due to each individual
term on the r.h.s. of Eq.~(\ref{poli}) is established, it is very easy to calculate the fine tuning reductions 
due to eventual cancellations. 
Let us assume, for example, that some soft parameters $p_i$ depend on one ``fundamental" parameter $p_0$ 
because of new physics beyond the GUT scale: $p_i=a_i p_0$. 
It is straightforward to see that the fine tuning due to $p_0$ 
is just the sum with signs of the individual outputs of $\textrm{FT}_i=p_i^2/M_Z^2\cdot\partial M_Z^2/\partial p_i^2$\,: 
\begin{equation}
\frac{\partial\ln M_Z^2}{\partial\ln p_0^2}=\frac{p_0^2}{M_Z^2}\sum_i\left(a_i^2\frac{\partial M_Z^2}{\partial p_i^2}\right)=\sum_i \textrm{FT}_i\,.\label{totFT}
\end{equation}

\section{Fine tuning of the MSSM with GUT constraints}\label{sec:gut_fine}

In this section we analyze  
the fine tuning of some popular GUT-constrained models, accounting for the effects of cancellations
induced by GUT-scale relations. We start in \refsec{sec:CMSSM} with the CMSSM,
which, as the simplest model of the class,  we use as 
a sort of ``meter stick" to measure the reduction of fine tuning when certain non-universal GUT-scale
assumptions are considered, as discussed in \refsec{sec:NUGM}. 
In \refsec{sec:NUscan} we will quantify the fine tuning reduction due to these relations.
   
The likelihood $\mathcal{L}$ for a point in the parameter space is evaluated using the \chisq\ statistics as a sum of individual contributions from 
the experimental constraints listed in Table~\ref{tab:exp_constraints}. Confidence regions are calculated 
with the profile-likelihood method from tabulated values
of $\delta\chisq\equiv-2\ln(\mathcal{L}/\mathcal{L}_{\textrm{max}})$.
In two dimensions, 68.3\% confidence regions are given
by $\delta\chi^2=2.30$ and 95.0\% confidence regions by $\delta\chi^2=5.99$.  
Throughout this study, we present points belonging to the 95\%~($\sim2\sigma$)
confidence regions in two-dimensional projections of the profile-likelihood.   
   
\begin{table}[t]\footnotesize
\begin{center}
\begin{tabular}{|l|l|l|l|l|l|}
\hline
Measurement & Mean or range & Error:~exp.,~th. & Distribution & Ref.\\
\hline
\mhl\ (by CMS) & $125.7\gev$ & $0.4\gev, 3\gev$ & Gaussian &\cite{CMS-PAS-HIG-13-005} \\
\abundchi 			& $0.1199$ 	& $0.0027$,~$10\%$ 		& Gaussian &  \cite{Ade:2013zuv}\\
\brbxsgamma $\times 10^{4}$ 		& $3.43$   	& $0.22$,~$0.21$ 		& Gaussian &  \cite{bsgamma}\\
\brbutaunu $\times 10^{4}$          & $0.72$  	& $0.27$,~$0.38$ 		& Gaussian &  \cite{Adachi:2012mm}\\
$\Delta M_{B_s}$ & $17.719\ps^{-1}$ & $0.043\ps^{-1},~2.400\ps^{-1}$ & Gaussian & \cite{Beringer:1900zz}\\
\sinsqeff 			& $0.23146$     & $0.00012$, $0.00015$             & Gaussian &  \cite{Beringer:1900zz}\\
$M_W$                     	& $80.385\gev$      & $0.015\gev$, $0.015\gev$               & Gaussian &  \cite{Beringer:1900zz}\\
$\brbsmumu\times 10^9$			& 2.9
&  0.7, 10\% & Gaussian &  \cite{Aaij:2013aka,Chatrchyan:2013bka}\\
\mbmbmsbar\ & 4.18\gev\ & 0.03\gev, 0 & Gaussian & \cite{Beringer:1900zz} \\
\mtpole\ & 173.5\gev & 1.0\gev, 0 & Gaussian & \cite{Beringer:1900zz} \\
LUX (2013) & See text. 	& See text. 	& Poisson &\cite{Akerib:2013tjd}\\ 
\hline
\end{tabular}

\caption{\footnotesize
The experimental constraints that we include in the likelihood function to constrain our models.
} 
\label{tab:exp_constraints}
\end{center}
\end{table}

In all the cases in our numerical scans we include the experimental constraints from the Higgs mass and signal rates, 
the relic density, EW and flavor physics. They are imposed through the likelihood function, as explained above.
The numerical analysis was performed with the package BayesFITSv3.1,
described in detail in\cite{Fowlie:2012im,Kowalska:2012gs,Fowlie:2013oua}. 
The package is linked to \texttt{\multinest\ v2.7}\cite{Feroz:2008xx}
for sampling. Mass spectra and fine tuning are calculated with \texttt{\softsusy\ v3.3.9}\cite{softsusy};
the branching ratios \brbxsgamma\ and \brbsmumu\ with \texttt{\superiso\ v3.3}\cite{superiso};
the relic density and spin-independent neutralino-proton cross section \sigsip\ 
with \texttt{\micromegas\ v3.2}\cite{micromegas}; and 
EW precision constraints with  \texttt{\feynhiggs\ v2.9.4}\cite{feynhiggs:99,feynhiggs:00,feynhiggs:03,feynhiggs:06}.
To include the exclusion limits from Higgs boson searches at LEP, Tevatron, and the LHC we use
\texttt{\higgsbounds\ v3.8.1}\cite{Bechtle:2008jh,Bechtle:2011sb,Bechtle:2013wla}, while the \chisq\ contributions from the Higgs boson signal rates
from Tevatron and the LHC are calculated with \texttt{\higgssignals\ v1.0.0}\cite{Bechtle:2013xfa}.

The likelihood relative to the recent LUX results\cite{Akerib:2013tjd}
was calculated by closely following the procedure for the XENON100 likelihood developed in\cite{Cheung:2012xb}
and updated in Sec.~3.2 of\cite{Fowlie:2013oua}.
We assume that the number of observed events, $o$, follows a Poisson distribution 
about the number of signal+background events, $s+b$,
\begin{equation}
\mathcal{P}(s+b|o)= \int_0^\infty\frac{e^{-(s + b')}\left(s + b' \right)^{o}}{o !}
\exp \left[ -\frac{(b'-b)^2}{2\delta b^2}\right]db'\,.\label{bgmarg}
\end{equation}
As can be seen in Eq.~(\ref{bgmarg}), the systematic uncertainties are accounted for by marginalizing
the background prediction with a Gaussian distribution.
We use the values given in\cite{Akerib:2013tjd}: $o=3.1\textrm{ mDRU}_{\textrm{ee}}$ ($1\textrm{ mDRU}_{\textrm{ee}}$ 
is $10^{-3}$~events/KeV$_{\textrm{ee}}$/kg/day), $b=2.6\textrm{ mDRU}_{\textrm{ee}}$,
$\delta b=\sqrt{2\cdot 0.2^2+0.4^2}$ $\textrm{ mDRU}_{\textrm{ee}}$, and the signal is simulated with   
\texttt{\micromegas} in the nuclear recoil energy range of 3--25~keV$_{\textrm{nr}}$.

Note that, in practice, the constraints on the Higgs mass and \abundchi\ are by far the strongest and play the most important
role in determining the allowed parameter space. In particular, in the $\sim 1\tev$ higgsino region the other 
constraints are easily satisfied as SUSY particles are very heavy and their contributions to the observables
are suppressed.
In some regions characterized by neutralinos with a mixed bino/higgsino composition the LUX likelihood 
plays a significant role and we will point out in the text when this is the case.

\subsection{CMSSM}\label{sec:CMSSM} 

As was explained in \refsec{intro:sec}, SUSY models defined 
in terms of high-scale boundary conditions are in general characterized 
by large levels of fine tuning because $\Delta\sim\ln(M_X/\msusy)$. On the other hand, the induced relations 
among parameters can translate into regions of low fine tuning due to the focusing
mechanism, as is the case of the FP region\cite{Chan:1997bi,Feng:1999mn,Feng:1999zg,Feng:2011aa,Feng:2012jfa} of the CMSSM.
Here we consider fine tuning in the CMSSM, which we use as 
a model of reference for the following cases.
   
In the CMSSM the fundamental GUT-scale parameters
are the unified scalar mass, \mzero, the unified gaugino mass, \mhalf, the unified trilinear parameter, \azero, the unified bilinear parameter, $B_0$, 
and the high-scale Higgs/higgsino mass parameter, $\mu_0$. 

To obtain an approximate estimate of the impact of the parameters on the
parameter space, one can recast Eq.~(\ref{poli}) as
\be\label{poli_cmssm}
\mhusq(\msusy)=0.074 m_0^2-1.008\mhalf^2-0.080 \azero^2+0.406\mhalf\azero\,.
\ee
The coefficient multiplying $m_0^2$ is the smallest, 
resulting in general in low scalar fine tuning, with a consequently low total fine tuning 
in the regions where 
\mzero\ is of the order of a few \tev\ but $\mu$, \azero, and \mhalf\ are not too large (the FP region).
However, the focusing in the scalar sector loses its efficiency with increasing \mzero. 
One finds $\dmzero\simeq20$ 
for $\mzero=1\tev$, but $\dmzero\simeq 500$ for $\mzero=5\tev$.

We scanned the CMSSM parameter space in the following  
broad ranges for \mzero, \mhalf:
\begin{eqnarray}
0.1\tev\leq&\mzero&\leq 10\tev\,,\nonumber\\
0.1\tev\leq&\mhalf&\leq 4\tev\,.\label{ranges1}
\end{eqnarray}
In order to minimize the impact of \azero\ and \tanb\ 
on the total fine tuning ($B\mu$, as usual, is traded for \tanb) 
we scanned those parameters in the following limited ranges:
\begin{eqnarray}
-1\tev\leq&\azero&\leq 1\tev\,,\nonumber\\
10\leq&\tanb&\leq 62\,.\label{ranges2}
\end{eqnarray}
The choice of a limited range for \azero\ and \tanb\
does not affect significantly the distribution of the profile likelihood
in the (\mzero, \mhalf) plane, with the exception of the stau-coannihilation region\cite{Ellis:1998kh},
which is not allowed in the ranges of (\ref{ranges2}) because it requires large mixing between the stops in order to obtain 
the right value of the Higgs mass\cite{Fowlie:2012im,Kowalska:2013hha}.
It is known, however, that the stau-coannihilation region of the CMSSM presents large 
values for $\Delta_{A_0}$, so that we do not treat it in this paper.  

\begin{figure}[t]
\centering
\subfloat[]{%
\label{fig:a}%
\includegraphics[width=0.50\textwidth]{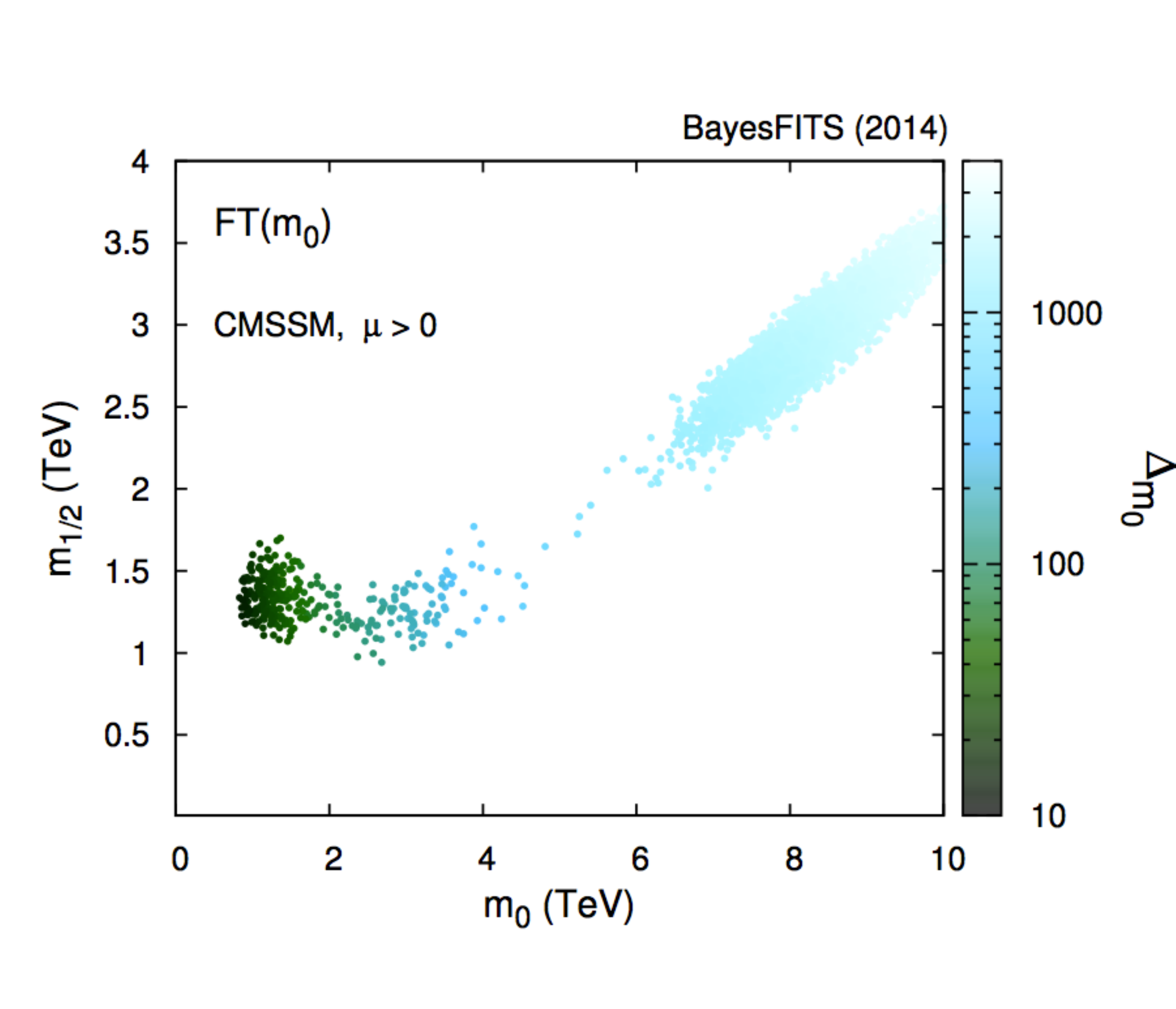}
}
\subfloat[]{%
\label{fig:b}%
\includegraphics[width=0.50\textwidth]{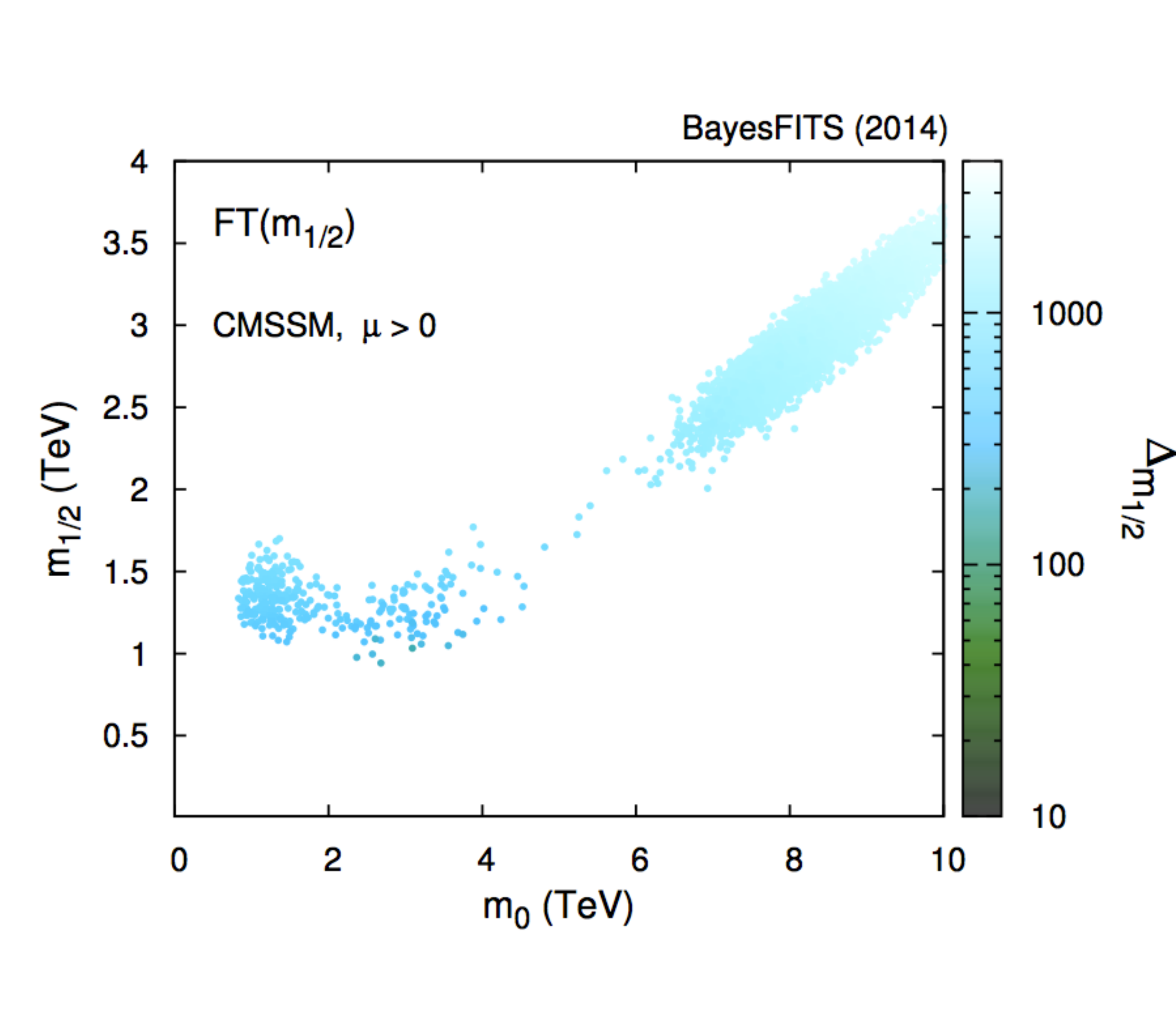}
}%
\vspace{-0.5cm}
\subfloat[]{%
\label{fig:c}%
\includegraphics[width=0.50\textwidth]{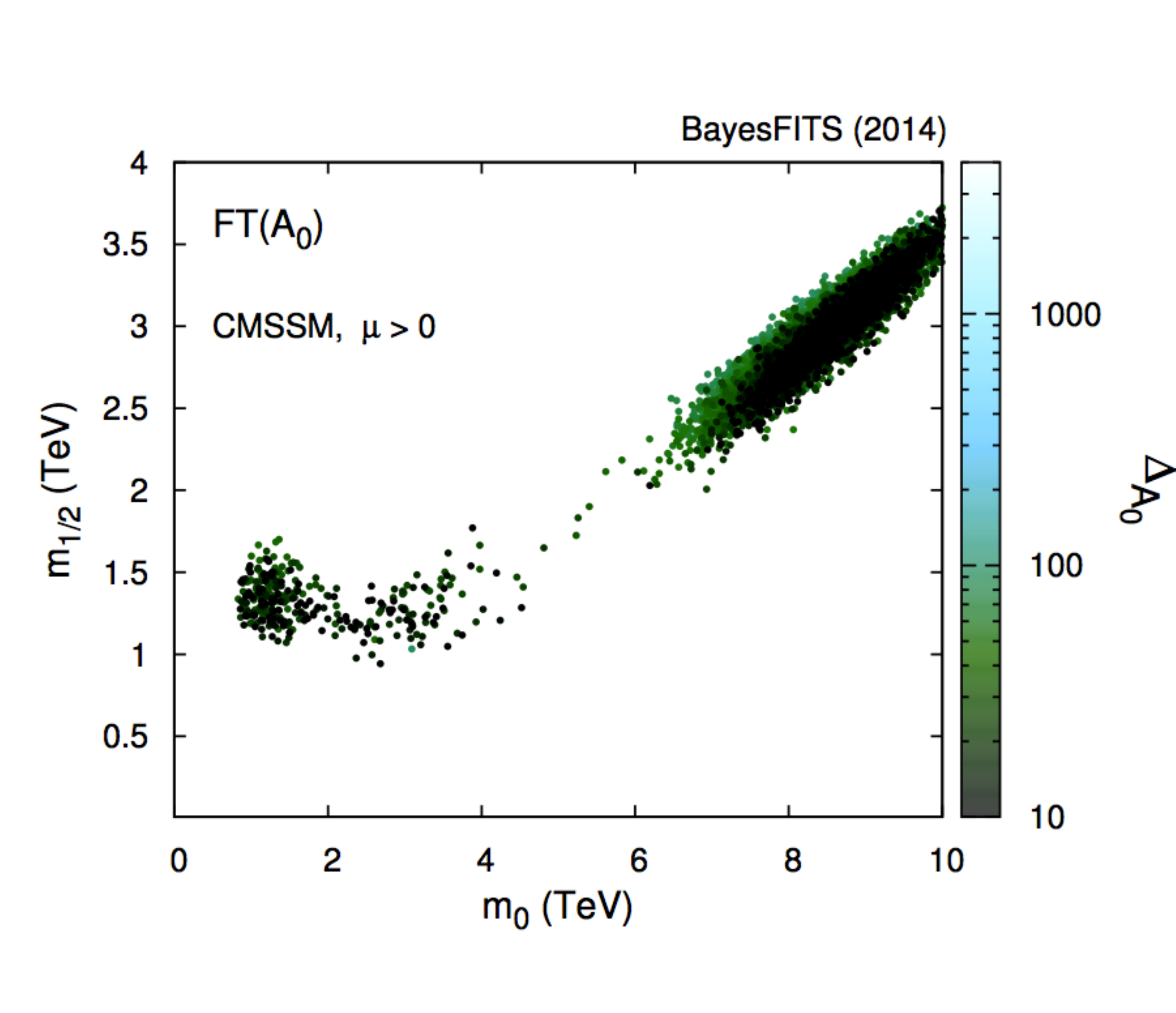}
}
\subfloat[]{%
\label{fig:d}%
\includegraphics[width=0.50\textwidth]{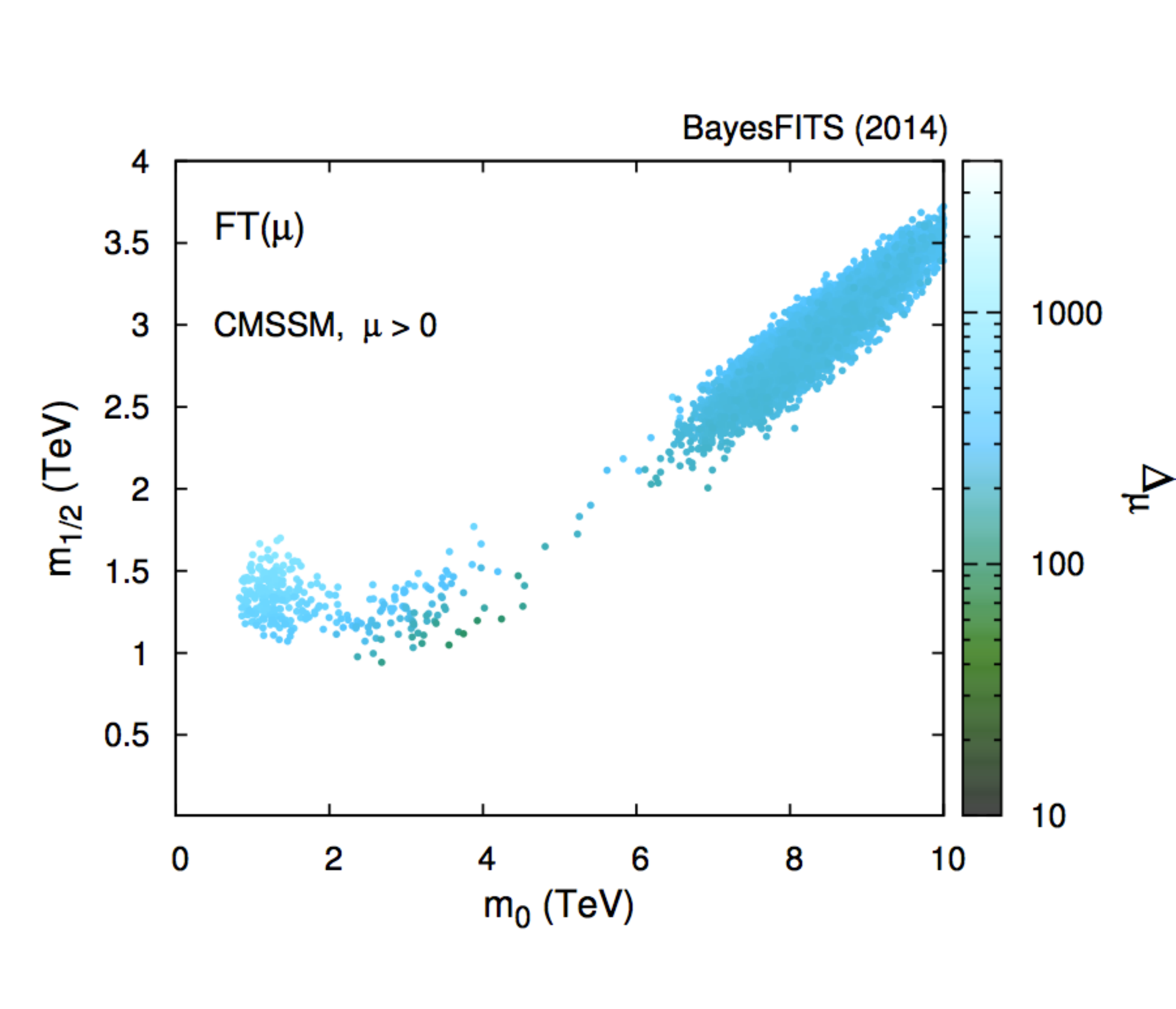}
}%
\caption{\footnotesize Scatter plots of the fine-tuning measure due to the different input parameters of the CMSSM.
All the points satisfy the constraints of Table~\ref{tab:exp_constraints} at $2\sigma$. \protect\subref{fig:a} $\Delta_{\mzero}$, 
\protect\subref{fig:b} $\Delta_{\mhalf}$, \protect\subref{fig:c} $\Delta_{A_0}$, and \protect\subref{fig:d} $\Delta_{\mu}$.}
\label{fig:cmssm_ft}
\end{figure}

We show in \reffig{fig:cmssm_ft} the distribution in the (\mzero, \mhalf)
plane of the fine tuning contributions due to \subref{fig:a} \mzero, \subref{fig:b} \mhalf, \subref{fig:c} \azero, 
and \subref{fig:d} $\mu_0$. All the points satisfy the constraints 
of Table~\ref{tab:exp_constraints} at $2\sigma$. 
Due to the choice (\ref{ranges2}) of \tanb\ ranges, the values of $\Delta_{B\mu}$
are below 10 over the whole parameter space and we do not show its distribution.

A few features are immediately visible in \reffig{fig:cmssm_ft}:
in the region of $\mzero\lesssim4\tev$ the dominant contribution to the fine tuning is given by $\mu_0$,
$\Delta_{\mu}\sim 500-1000$,
with the exception of a few points at $\mzero\simeq 3-4\tev$
and $\mhalf\simeq 1\tev$, for which $\Delta_{\mu}\lesssim 100$
and $\Delta_{\mhalf}$ is dominant. Those are the points
adjacent to the FP region, where $\mu$ is lower.
The FP region appears to be nearly excluded
in the plots because it is disfavored by the LUX
likelihood. 
Note, however, that this tension can be ameliorated if one includes the theoretical uncertainties 
due to the nuclear physics $\Sigma_{\pi N}$ terms in the likelihood function\cite{Buchmueller:2013rsa}.

In the upper right part of the (\mzero, \mhalf) plane,
a very large region characterized by
a nearly pure higgsino LSP with $\mchi\simeq 1\tev$ is present. As discussed in \refsec{intro:sec},
in this region the relic abundance assumes the correct value and it is also most naturally 
compatible with $\mhl\simeq 126\gev$ due to large \msusy.
The fine tuning due to $\mu_0$, $\Delta_{\mu}\simeq 250$, is large but insensitive to varying the CMSSM parameters.  
The total fine tuning $\Delta$ is dominated by the contributions of multi-\tev\ scalar and gaugino masses.
Note, finally, that \reffig{fig:cmssm_ft}\subref{fig:d} shows that our choice
of the \azero\ range helps maintaining $\Delta_{\azero}$ well below 100 over 
large fractions of the parameter space.

\begin{figure}[t]
\centering
\includegraphics[width=0.60\textwidth]{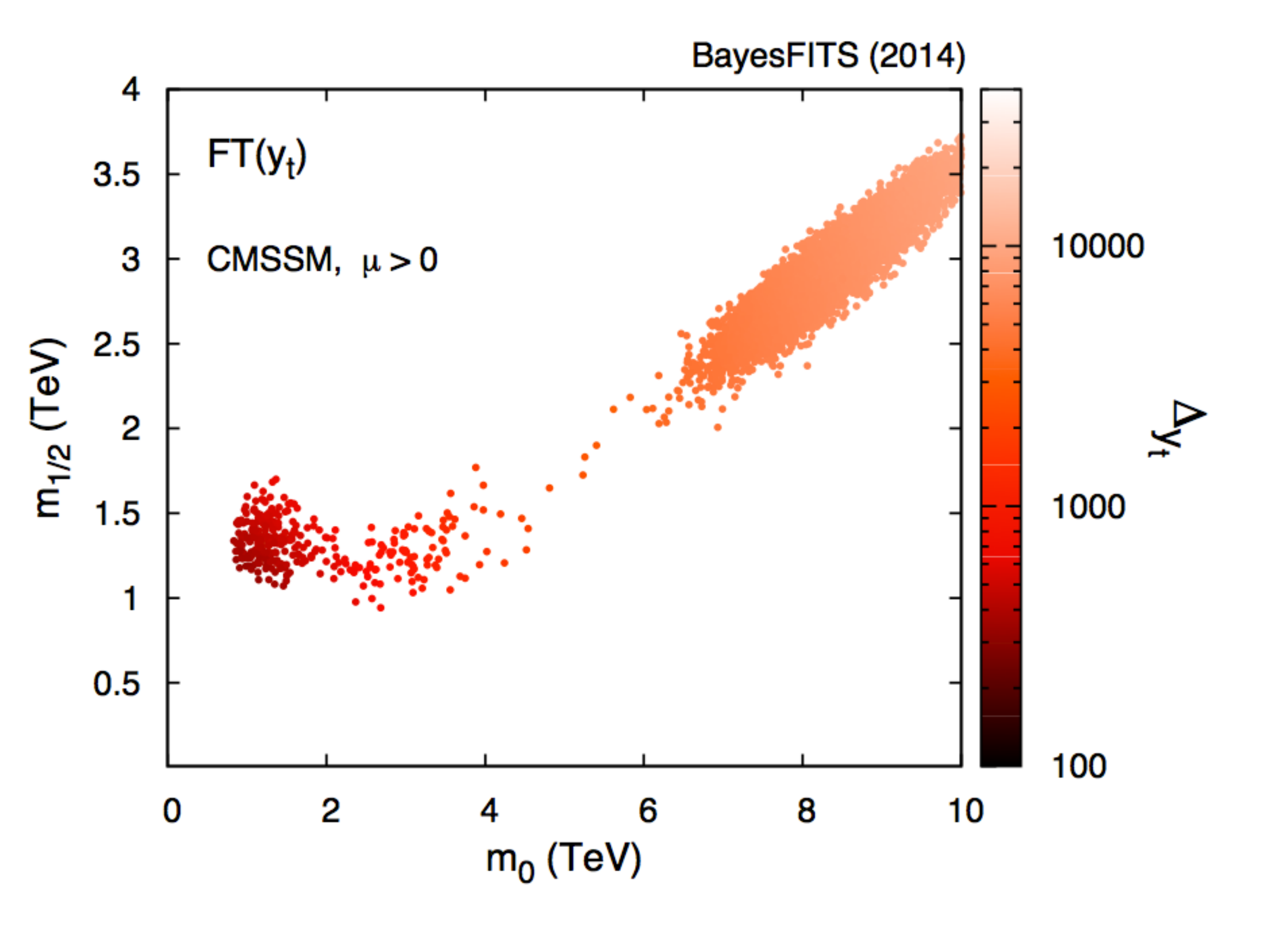}
\caption{\footnotesize Scatter plot of the fine-tuning measure due to the top Yukawa coupling in the (\mzero, \mhalf) plane of the CMSSM.
All the points satisfy the constraints of Table~\ref{tab:exp_constraints} at $2\sigma$.}
\label{fig:cmssm_yt}
\end{figure}

For completeness, we show in \reffig{fig:cmssm_yt} the distribution of the top-Yukawa fine tuning, $\Delta_{y_t}$, in the allowed regions of the CMSSM.
As was to be expected, since $\Delta\sim y_t^2 (\msusy/M_Z)^2$ $\ln(M_X/\msusy)$, when the scalar and gaugino masses become large
$\Delta_{y_t}$ can assume values of the order of $10^4$, thus becoming the dominant source of fine tuning.

The main goal of this study is to suitably choose the GUT-scale mass relations of the scalars and gauginos 
to reduce the overall fine tuning of the SUSY sector with respect to the CMSSM.
This is done by extending the mechanism of parameter focusing.  
Expansions of the form of (\ref{poli}) are obtained by integrating the RGEs for the SUSY-breaking parameters 
and fixing the initial conditions for the gauge and Yukawa couplings at the experimentally determined values. 
One can thus express 
the dependence of the EW scale on the values of the SUSY-breaking parameters at the high scale. 
As will be clear in \refsec{sec:NUGM} and the following sections,
the relations leading to the lowest levels of fine tuning are those involving parameters characterized by the largest coefficients in (\ref{poli}).
Searching for such mass relations assumes specific initial values of the gauge and Yukawa couplings, and 
including the latter in the analysis would make finding efficient focusing mechanisms much more complicated,
especially if one wanted to take into account relations suggested by frameworks of family symmetries. 
Thus, in this paper we have decided to concentrate specifically on the fine tuning of the SUSY sector of the theory. 

\subsection{Non-universal gaugino and scalar masses}\label{sec:NUGM} 

As was shown in \refsec{sec:CMSSM}, in the broad $\sim 1\tev$ higgsino region the fine tuning is dominated by the 
contributions of large scalar and gaugino masses, 
while the contribution due to $\mu_0$ is smaller and
stable over the whole region.

The fine tuning in the gaugino sector can be significantly ameliorated in models 
in which GUT-scale universality is abandoned, i.e. $M_1\neq M_2\neq M_3$ at the GUT scale.
Particularly it was shown\cite{Kane:1998im,BasteroGil:1999gu,Kane:2002ap,Abe:2007kf,Horton:2009ed,Antusch:2012gv,Gogoladze:2012yf,Gogoladze:2013wva} that the ratio $M_2/M_3\simeq 2.5-3$ 
leads to a significant reduction of the gaugino contribution to the total fine tuning with respect to the case with universality.
This can be easily understood by comparing the coefficients of $M_2^2$ and $M_3^2$ in Eq.~(\ref{poli}).

NUGM patterns can be generated naturally in the 
framework of supergravity embedded in some GUT gauge group\cite{Ellis:1985jn,Anderson:1996bg,Ellis:1984bm,Drees:1985bx,Giudice:1998xp}. 
The source of gaugino masses are non-renormalizable dimension-5 operators generated by a chiral superfield whose $F$-term acquires a vev. 
The field can belong to any irreducible representation in the symmetric product of two adjoints of the gauge group, 
leading to various fixed mass relations among the gaugino masses. In $SU(5)$\cite{Georgi:1974sy} 
there are four different patterns allowed\cite{Ellis:1985jn,Anderson:1996bg}:
\be
(M_1:M_2:M_3)\quad=\quad (1:1:1),\quad (10:2:1),\quad (-5:3:1),\quad (-\frac{1}{2}:-\frac{3}{2}:1)\,.\label{su5}
\ee
The corresponding ratios in $SO(10)$ have been obtained and summarized in\cite{Martin:2009ad}. 

\begin{figure}[t]
\centering
\subfloat[]{%
\label{fig:a}%
\includegraphics[width=0.50\textwidth]{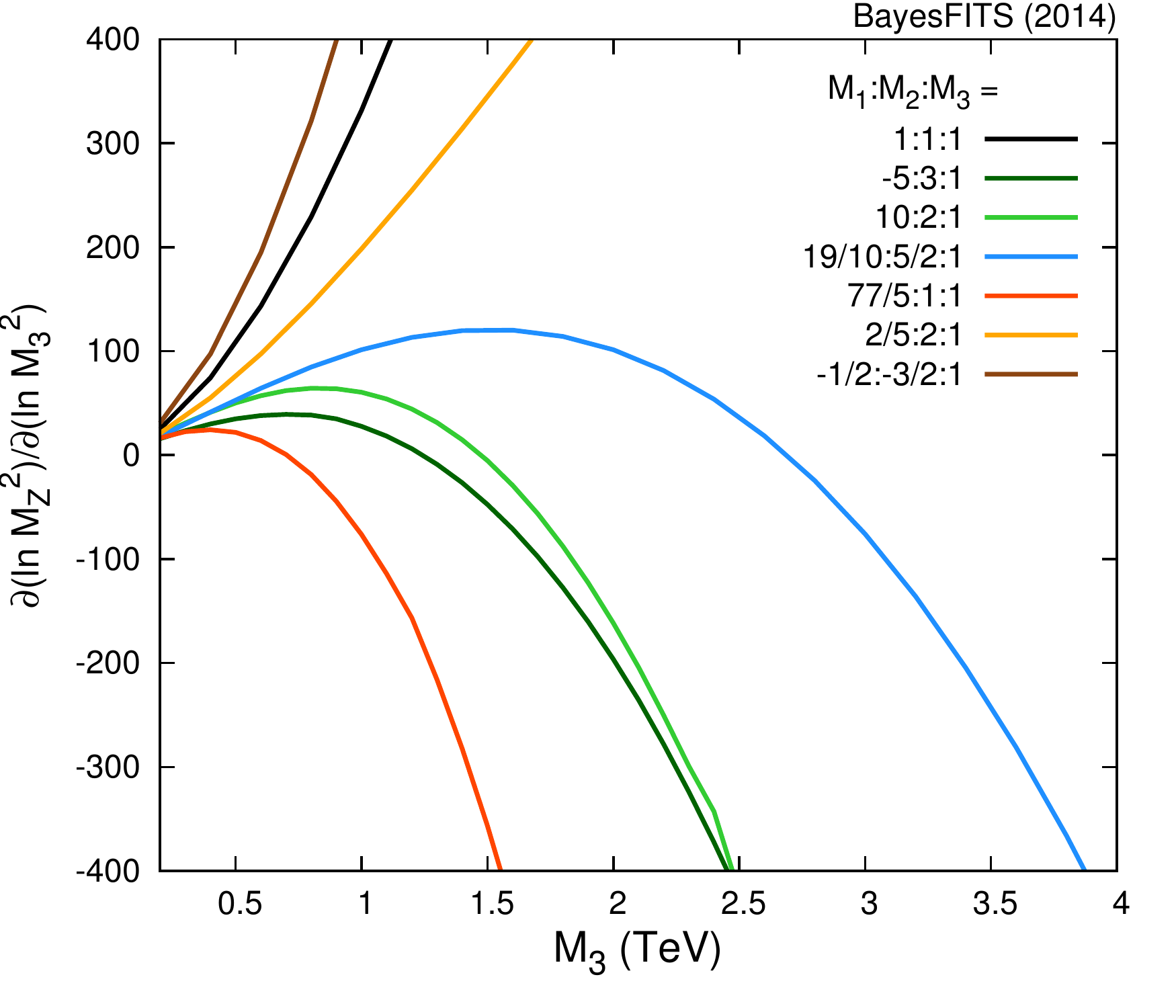}
}%
\subfloat[]{%
\label{fig:b}%
\includegraphics[width=0.50\textwidth]{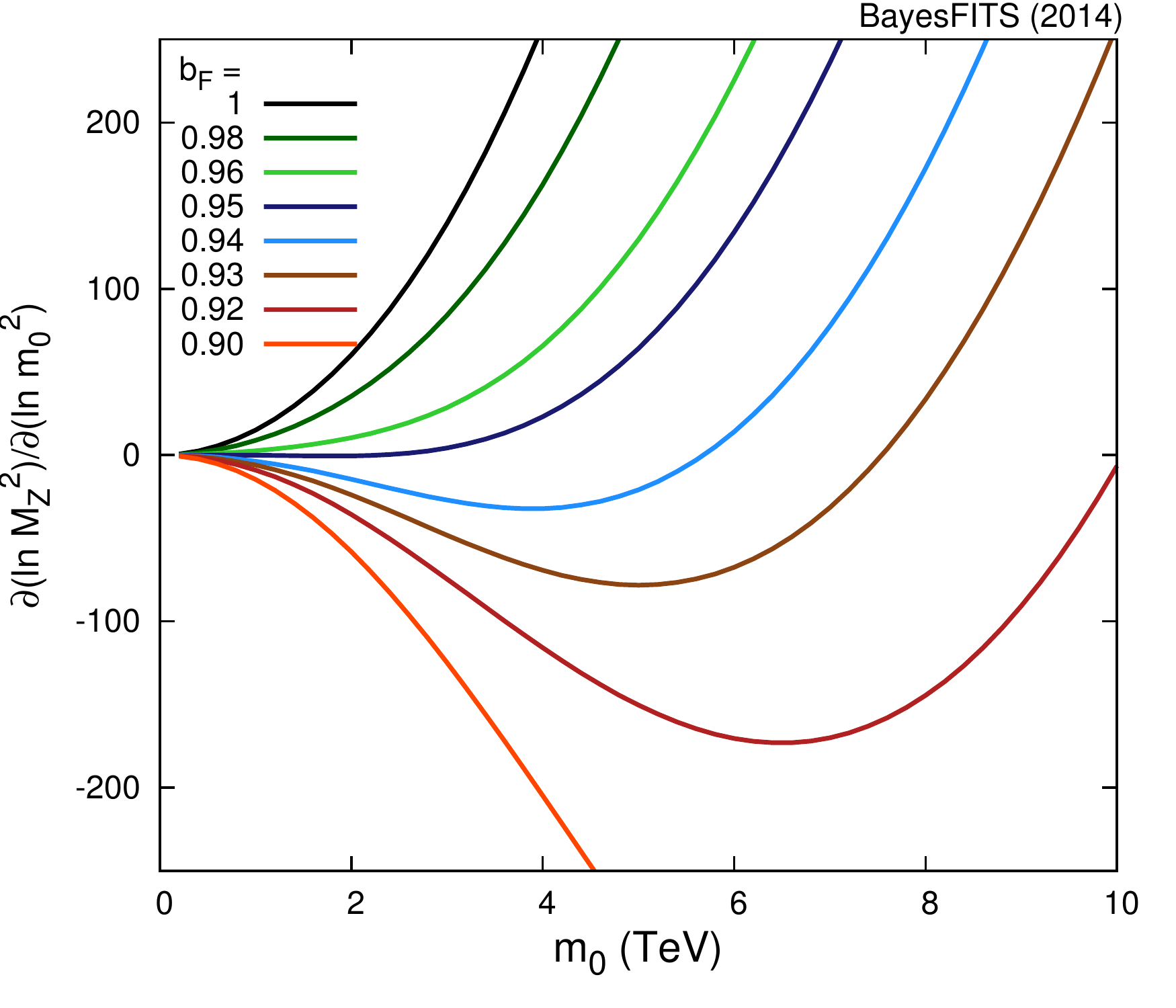}
}%
\caption{\footnotesize \protect\subref{fig:a} The fine tuning due to $M_3$ for different GUT-scale gaugino mass patterns.
$(10:2:1)$, $(-5:3:1)$, and $(-1/2:-3/2:1)$ come from representations of $SU(5)$\cite{Ellis:1985jn,Anderson:1996bg}.
$(19/10:5/2:1)$, $(77/5:1:1)$, and $(2/5:2:1)$ are some representative $SO(10)$ patterns\cite{Martin:2009ad}.
\protect\subref{fig:b} The fine tuning due to the unified scalar mass \mzero\ 
for different choices of the parameter $b_F=\mhu(\mgut)/\mzero$.}
\label{fig:ft_gau}
\end{figure}

Clearly, not all known gaugino mass patterns lead to a reduction of the 
gaugino fine tuning with respect to the universal case. 
This can be seen in \reffig{fig:ft_gau}\subref{fig:a}, where we plot the gaugino fine tuning calculated for the 
non-universal patterns (\ref{su5}) of $SU(5)$ and some of the patterns that can be generated in $SO(10)$
(we selected $M_3$ as the fundamental parameter in the gaugino sector). 
The curves are drawn for fixed $\mzero=1\tev$, $\azero=-1\tev$, and $\tanb=30$, but the features are generic.
As expected, the patterns ($19/10:5/2:1$), ($10:2:1$), and ($-5:3:1$) give the lowest levels of gaugino fine tuning and, 
most importantly, they do so over a large range of $M_3$ values.

One can also see in \reffig{fig:ft_gau}\subref{fig:a} that
in the ($19/10:5/2:1$) pattern $\partial \ln M_Z^2/\partial \ln M_3^2$ 
crosses zero for a relatively large value of $M_3$: $M_3\simeq 2700\gev$.
We find that a general feature of most gaugino mass patterns is the presence
of a region of large $M_3$ for which $\partial M_Z^2/\partial M_3^2\simeq0$ and, consequently,
$\Delta_{M_3}\simeq0$. However, \reffig{fig:ft_gau}\subref{fig:a} also shows that,
while the ($10:2:1$) and ($-5:3:1$) patterns present the same feature,
the distribution of fine tuning is
somewhat smoother, which can be explained mathematically with the fact
that overall $|\partial^2 M_Z^2/\partial^2 M_3^2|$ is smaller than in the 
($19/10:5/2:1$) case. 
We think that a smaller second derivative is an indication 
of better and less accidental stability of the EW scale, 
so that we will select the 
pattern ($-5:3:1$) as our case of choice.

In the scalar sector, the amount of fine tuning strongly depends on the assumed boundary conditions at the GUT scale, 
particularly on the high-scale relation among $\mhusq$, $\mqthree$, and $\muthree$\cite{Antusch:2012gv}. 
In $SU(5)$ (or in $SO(10)$)
the fermions and the Higgs bosons belong to different representations, 
so that the corresponding soft-breaking masses are in general 
unrelated and the fine tuning can become very large. 
If supergravity-inspired universality conditions are imposed at the high scale, the 
fine tuning can be reduced to the CMSSM levels shown in \reffig{fig:cmssm_ft}\subref{fig:a}. 

By rewriting $\mhusq(\msusy)$ in Eq.~(\ref{poli}) in terms of the common scalar mass, 
$m_0^2$, and \mhusq\ at the GUT scale ($\mhu(\mgut)\ne\mzero$ in what follows) one can derive
\be
\mhusq(\msusy)\simeq -0.571 m_0^2+0.645\mhusq+\textrm{gaugino and trilinear contributions}\,.
\label{expans}
\ee
It is straightforward to see that one can obtain less fine tuning from the scalars than in the CMSSM 
when $\mhusq(\mgut)$ and $m_0^2$ are related as
\be
\mhusq=b_F^2 m_0^2,\quad \textrm{with}\quad |b_F|\simeq \sqrt{0.57/0.64}=0.94\,.
\label{Defbf}
\ee
For simplicity we will consider $b_F$ to be positive.
Equation~(\ref{expans}) is approximate (although it holds rather well over most of the parameter space), 
but it gives a good estimate of the values of $b_F$ that are necessary to reduce the fine tuning with respect to the CMSSM, 
even for masses in the multi-TeV regime. Note that, remarkably, $b_F$ does not deviate substantially from 1, the value corresponding to universal scalar masses.

In \reffig{fig:ft_gau}\subref{fig:b}, we show the scalar fine tuning as a function of 
\mzero\ for different values of $b_F$. The curves are drawn for fixed values 
$\mhalf=1\tev$, $\azero=-1\tev$ and $\tanb=30$. 
Figure~\ref{fig:ft_gau}\subref{fig:b} also shows 
that values of $b_F\lesssim 0.93$ can produce low fine-tuning 
regions even with very large $\mzero$ values because at some point $\partial M_Z^2/\partial m_0^2\simeq 0$.
However, when $0.93\lesssim b_F\lesssim 0.94$ the region $m_0\lesssim8\tev$ features
consistent and stable values of low fine tuning, as $|\partial^2 M_Z^2/\partial^2 m_0^2|$
is generally smaller than for the other choices. 

In \refsec{sec:GMmech} we will comment on the possibility of 
generating non-universality in the scalar sector with supergravity. 
Alternatively it is possible to generate $b_F<1$ in the context of the MSSM embedded in a GUT symmetry 
and in \refsec{sec:model} we give an example of this for 
$SU(5)$.

\subsection{Non-universality and fine tuning in the allowed parameter space}\label{sec:NUscan}

Let us now turn to phenomenologically viable models and show how the conditions derived in \refsec{sec:NUGM} 
affect the fine tuning in the parameter space allowed by the constraints of Table~\ref{tab:exp_constraints}.

\begin{figure}[t]
\centering
\subfloat[]{
\label{fig:a}
\includegraphics[width=0.50\textwidth]{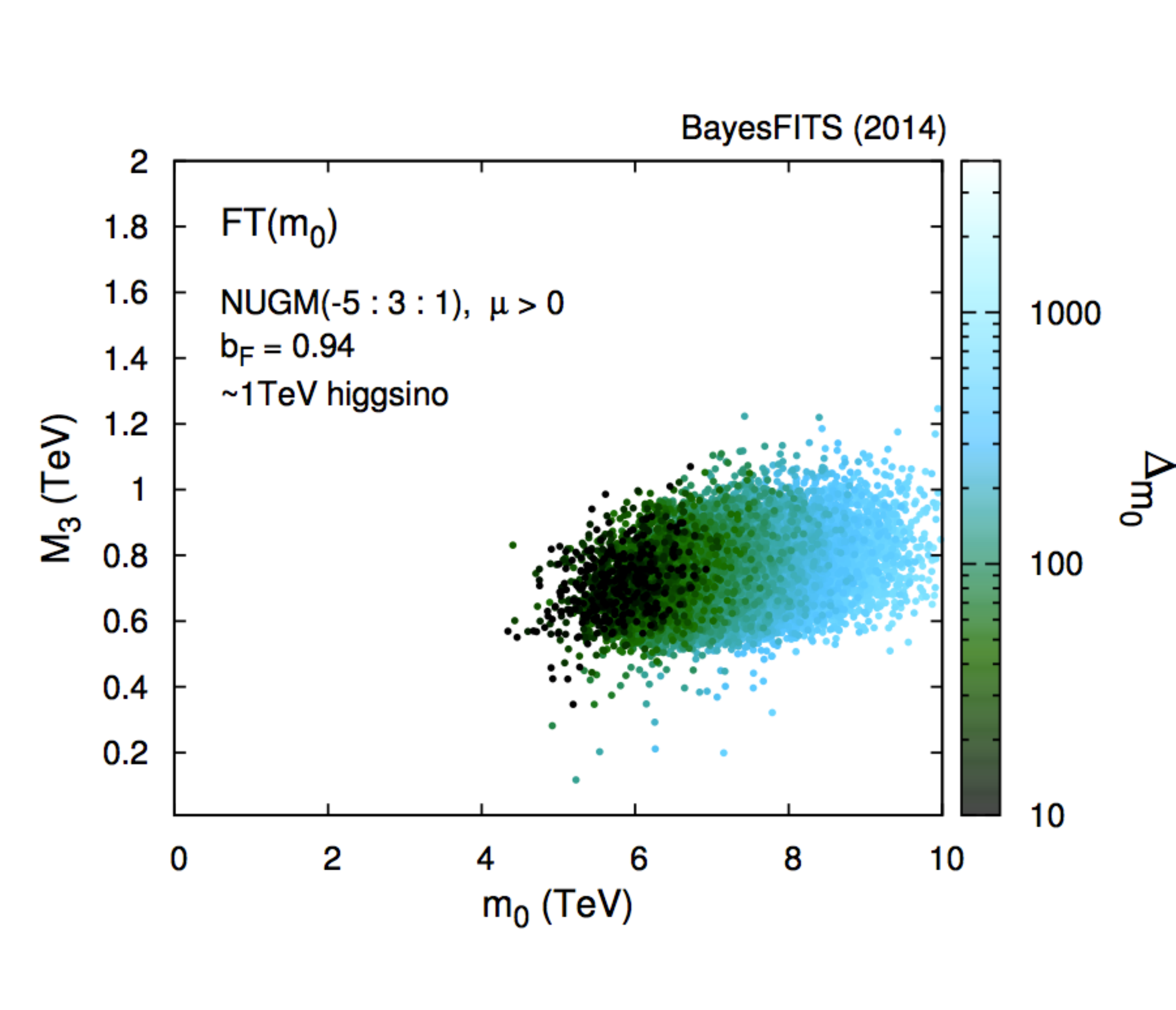}
}
\subfloat[]{
\label{fig:b}
\includegraphics[width=0.50\textwidth]{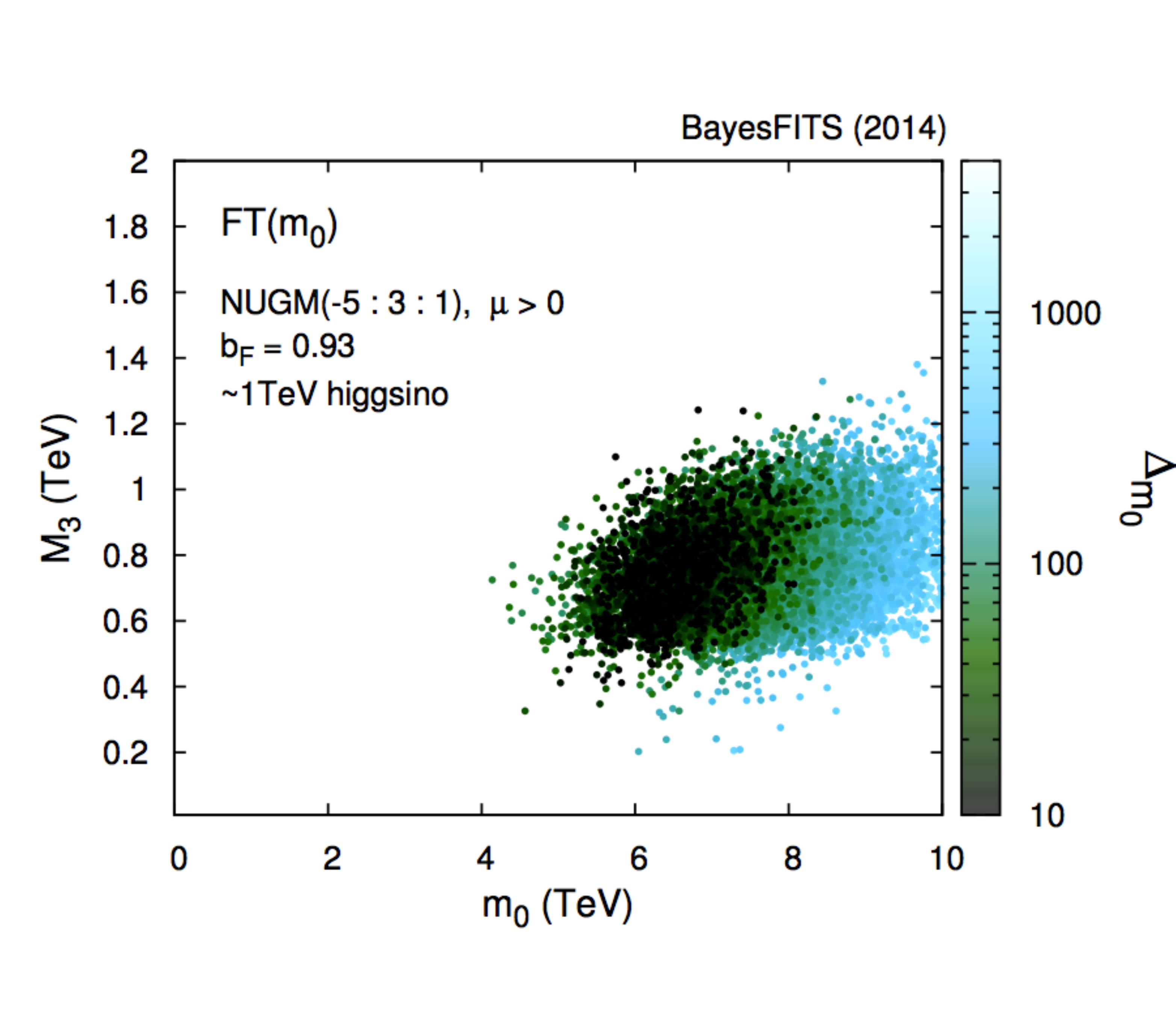}
}\\
\subfloat[]{
\label{fig:c}
\includegraphics[width=0.50\textwidth]{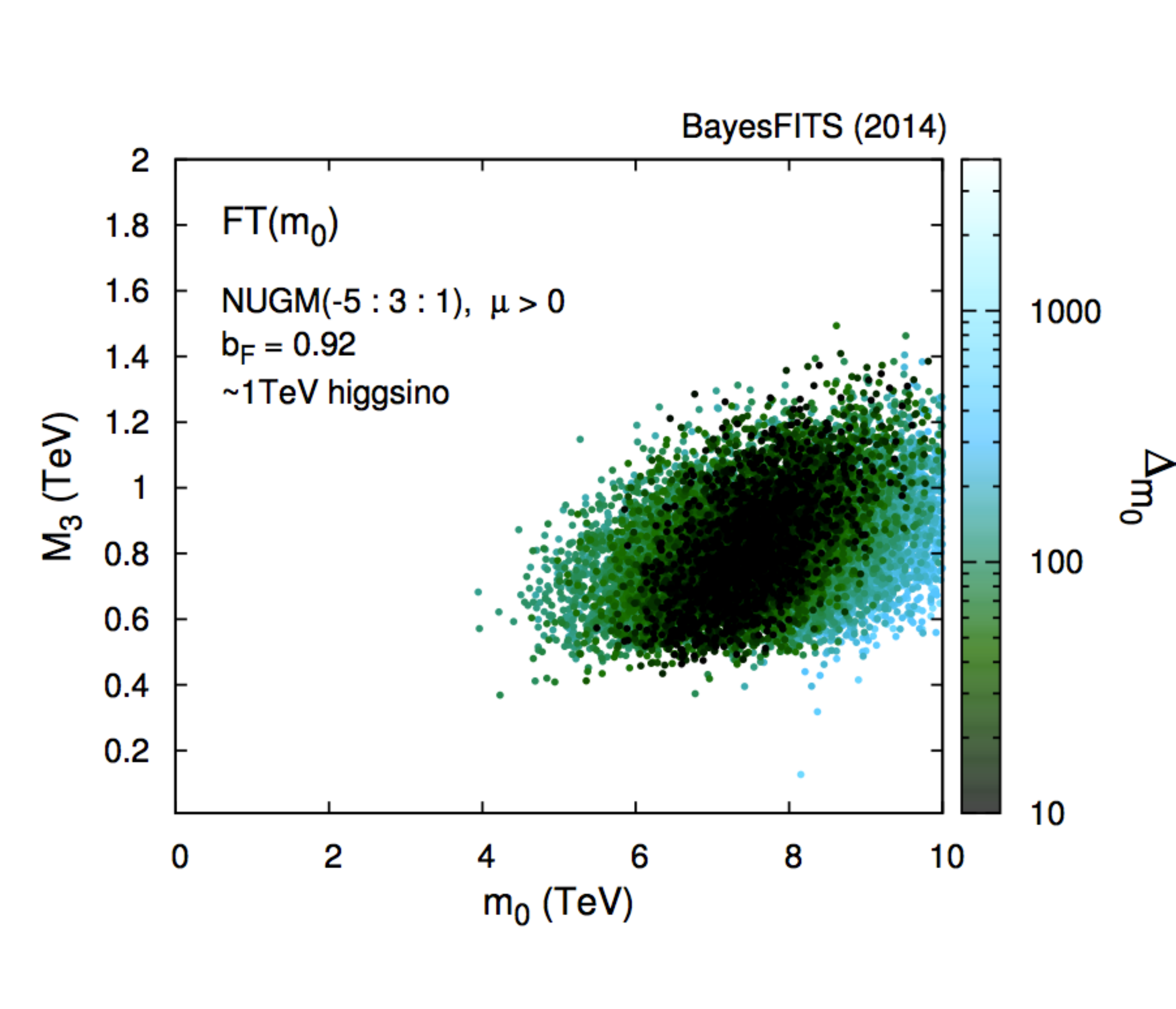}
}
\caption[]{\footnotesize Scatter plots of the fine tuning due to the scalars in the (\mzero, $M_3$) plane of NUGM ($-5:3:1$) 
for 3 different choices of $b_F=\mhu(\mgut)/\mzero$. \subref{fig:a} $b_F=0.94$ \subref{fig:b} $b_F=0.93$, and \subref{fig:c} $b_F=0.92$.
Note that it is in general lower than in the CMSSM, see \reffig{fig:cmssm_ft}\subref{fig:a}.
All points satisfy the constraints of Table~\ref{tab:exp_constraints} at $2\sigma$.}
\label{fig:NUGM2}
\end{figure}

With universal gaugino masses and $b_F<0.96$, the parameter ranges given by (\ref{ranges1}) and (\ref{ranges2})
do not always give the $\sim 1\tev$ higgsino LSP.  
In fact, by using Eqs.~(\ref{expans}) and (\ref{Defbf}) one can easily show that $\mhusq(\msusy)\simeq -1\tev^2$, 
which is typical of the $\sim 1\tev$ higgsino region, can only be obtained for much larger \mzero\ when $b_F<1$, 
as this effectively leads to a strong reduction of the coefficient multiplying $m_0^2$ with respect to (\ref{poli_cmssm}).

Conversely, if all GUT-scale scalar masses are kept equal, not all gaugino mass pattern can produce 
physical or phenomenologically acceptable regions. Some lead to
no EWSB over large parts of the parameter space, as is the case of the ($-5:3:1$) pattern;
others are physically viable but cannot produce regions that satisfy 
the constraints of Table~\ref{tab:exp_constraints}, as is the case of the ($10:2:1$) pattern in the ranges of (\ref{ranges1}) and (\ref{ranges2}).
 
Thus, non-universality conditions both in the scalar and gaugino sectors can not only be seen as instrumental to obtaining lower levels of fine tuning, 
but also as favoring phenomenologically acceptable regions extended over a large parameter space.

\begin{figure}[t]
\centering
\subfloat[]{
\label{fig:a}
\includegraphics[width=0.50\textwidth]{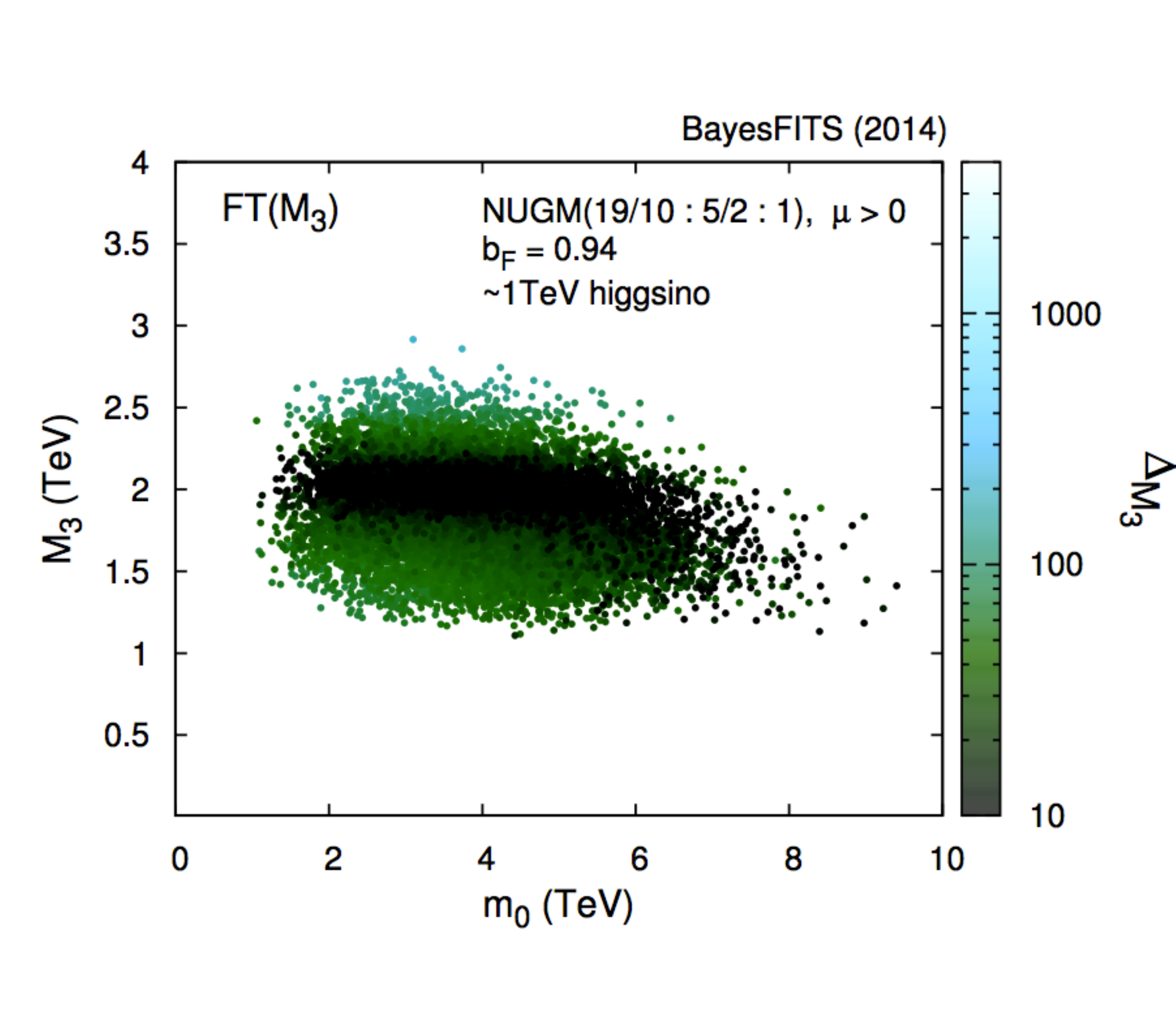}
}
\subfloat[]{
\label{fig:b}
\includegraphics[width=0.50\textwidth]{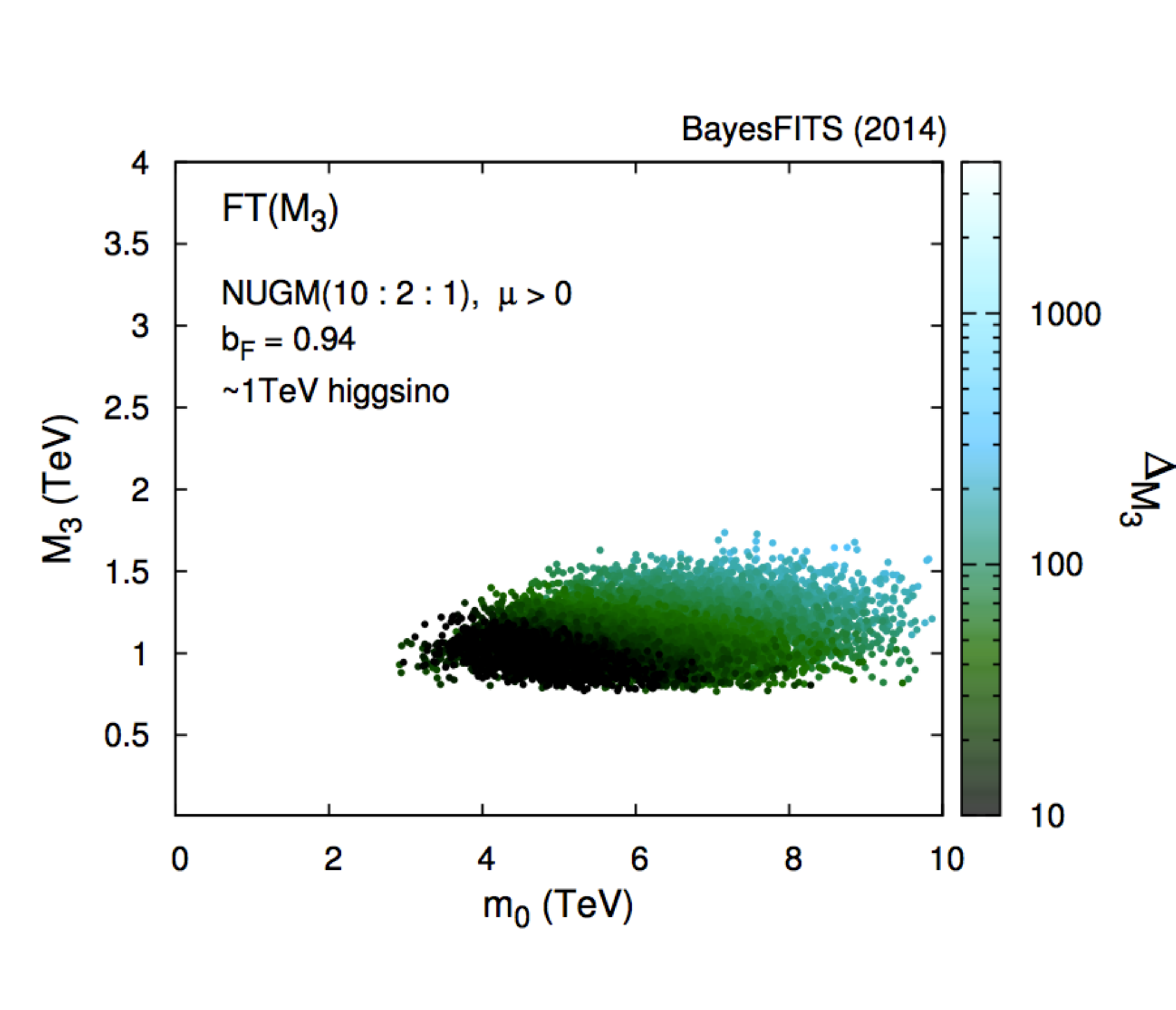}
}\\
\subfloat[]{
\label{fig:c}
\includegraphics[width=0.50\textwidth]{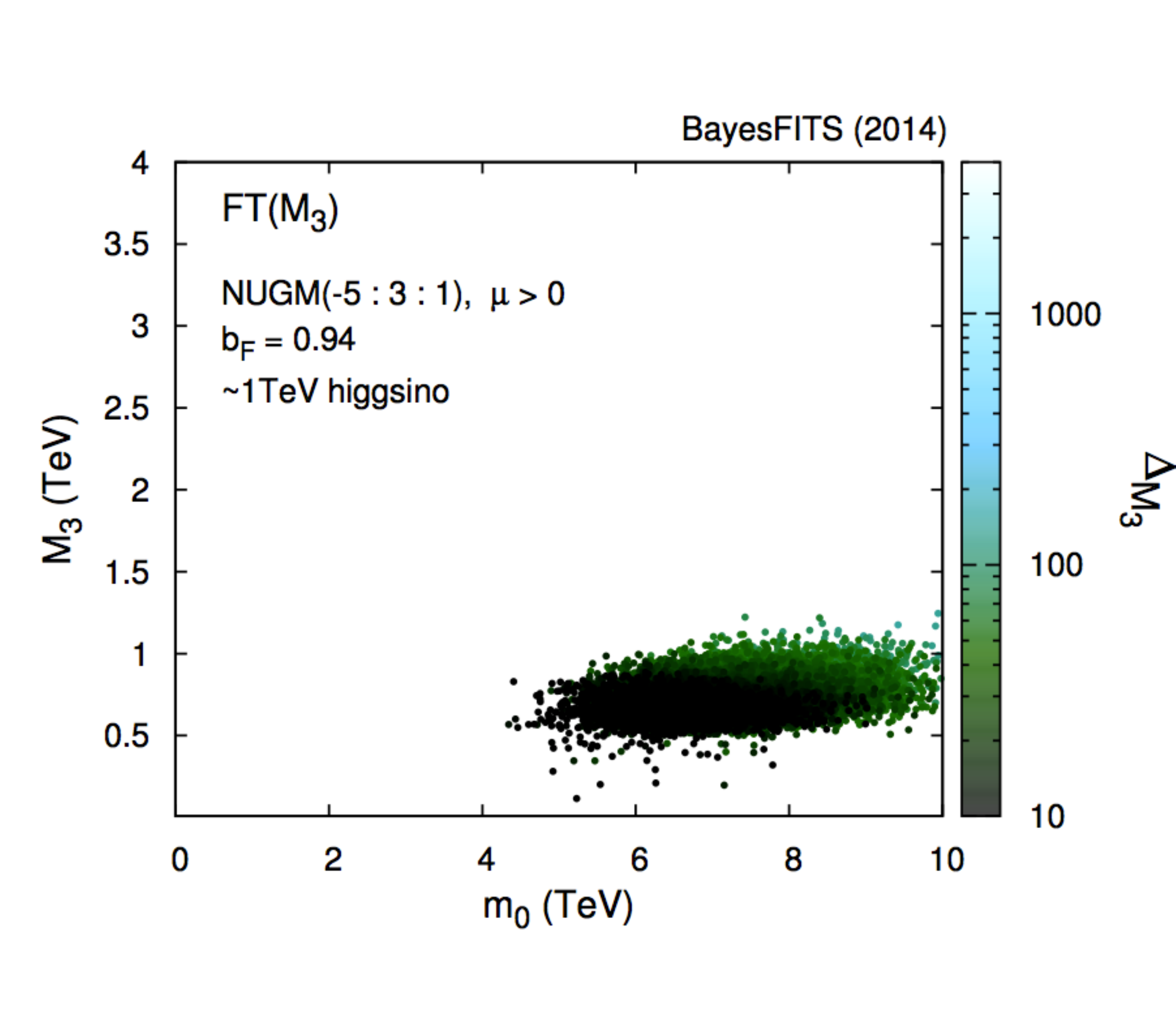}
}
\caption[]{\footnotesize Scatter plots of the gaugino fine tuning in the (\mzero, $M_3$) plane for 3 different gaugino mass patterns:
\subref{fig:a} ($19/10:5/2:1$), \subref{fig:b} ($10:2:1$), and \subref{fig:c} ($-5:3:1$). Note that it is in general lower than in the CMSSM, see \reffig{fig:cmssm_ft}\subref{fig:b}.
All points satisfy the constraints of Table~\ref{tab:exp_constraints} at $2\sigma$.}
\label{fig:NUGM}
\end{figure}

In \reffig{fig:NUGM2} we show the distribution of $\Delta_{\mzero}$ 
in the (\mzero, $M_3$) plane of a specific NUGM, ($-5:3:1$), for three different choices of $b_F<1$.
The parameter space was scanned in the ranges of (\ref{ranges1}) and (\ref{ranges2}), 
where \mhalf\ was replaced by $M_3$, and the additional assumption of $\mhu(\mgut)=b_F \mzero$
was adopted. In all the cases the points satisfying the constraints at $2\sigma$ belong to the $\sim 1\tev$ higgsino region.
One can see that the allowed region of parameter space moves towards larger \mzero\ with decreasing $b_F$, as explained above, 
while remaining very weakly dependent on $M_3$. 
The selected $b_F$ values consistently allow one to identify regions of the parameter space 
with $\Delta_{\mzero}\ll 10$. 

In \reffig{fig:NUGM} we show the distribution of $\Delta_{M_3}$ in the (\mzero, $M_3$) plane for the 
three gaugino mass patterns that give the lowest fine tuning in \reffig{fig:ft_gau}\subref{fig:a}: 
\subref{fig:a} ($19/10:5/2:1$), \subref{fig:b} ($10:2:1$), and \subref{fig:c} ($-5:3:1$).
Different mass relations between gauginos at the GUT scale can have different phenomenological implications, 
as the latter depend 
on the low-scale ratios between the bino, wino and higgsino masses (see, e.g.,\cite{Griest:1991gu}
for an early study). However, for any chosen gaugino pattern, the value of $b_F$ barely affects the distribution of gaugino fine tuning. 
We adopt, in \reffig{fig:NUGM}, $b_F=0.94$.
We scanned the parameter space over the ranges given in (\ref{ranges1}) and (\ref{ranges2}), 
where \mhalf\ was replaced by $M_3$.
Again, in the three selected cases the points satisfying the constraints at $2\sigma$ belong to the $\sim 1\tev$ higgsino region.

Interestingly, in all three cases $\Delta_{M_3}$ is consistently below 100
and can even reach values $\ll 10$ 
over broad regions of the allowed parameter space, in striking contrast with 
the case of gaugino universality (CMSSM, \reffig{fig:cmssm_ft}\subref{fig:b}),
for which $\Delta_{m_{1/2}}\gsim 1000$ in the $\sim1\tev$ higgsino region. 
In \reffig{fig:NUGM}\subref{fig:a} the region where the fine tuning is the lowest is restricted
to a horizontal strip about $M_3\simeq2\tev$.
This behavior was explained in \refsec{sec:NUGM} and it has to do with the fact that  
$\partial M_Z^2/\partial M_3^2\simeq0$\,.

\section{$\mu$ term}\label{sec:muterm}

We have shown in the previous section that, if one assumes certain relations among the input parameters resulting from high-scale physics, 
the mechanism of focusing induces significantly lower gaugino and scalar fine tuning than in the CMSSM.  

In the region of moderate-to-large \tanb\ and relatively small GUT-scale trilinear terms, the only remaining issue 
is the impact of the $\mu$ parameter. Without any further assumption, from the point of view of 
reducing fine tuning, $\mu$ should probably be as small as possible, see Eq.~(\ref{naturb}).
The phenomenology of this case has been extensively treated in the literature\cite{Baer:2012up,Baer:2012mv,Baer:2013xua,Baer:2013vpa,Schwaller:2013baa,Baer:2014cua}.
As was mentioned in \refsec{intro:sec}, ``low $\mu$" scenarios generally imply an under-abundance of dark matter,
as the LSP is a quite light higgsino-like neutralino. 
A higgsino LSP at $\sim1\tev$, on the other hand, does satisfy the constraints from PLANCK, but produces a
somewhat larger level of fine tuning.

Several mechanisms that can relate the $\mu$
parameter to the other soft terms have been known in the literature, the most popular probably being the one invented 
by Giudice and Masiero\cite{Giudice:1988yz} in supergravity, 
which relates $\mu_0$ to the scalar soft masses. 
We recall the main points of the mechanism in \refsec{sec:GMmech} 
and we show that the ensuing relation between $\mu_0$ and \mzero\
can be used to significantly reduce  
the overall level of fine tuning of the $\sim1\tev$ higgsino region.
We present some phenomenological properties of these low fine-tuning scenarios
in \refsec{sec:pheno}.
  
Alternative mechanisms that can similarly reduce the fine tuning by relating $\mu$ to  
other parameters through the EWSB conditions have been developed, e.g.,   
in the context of the NMSSM\cite{Ross:2011xv,Ross:2012nr}.

%


\subsection{New regions of low fine tuning from a $\mathbf{\mu_0\sim\mzero}$ relation.}\label{sec:GMmech}

The Giudice-Masiero mechanism\cite{Giudice:1988yz} gives an elegant framework to 
solve the $\mu$ problem of the MSSM\cite{Kim:1983dt}, in the context of local supersymmetry-breaking, or $N=1$ 
supergravity\cite{Hall:1983iz,Soni:1983rm}.

The minimal implementation entails a set of visible-sector 
superfields, $\hat{C}_i$, $\hat{C}_i^{\dag}$, and at least one set of hidden-sector superfields,
$\hat{h}$, $\hat{h}^{\dag}$. If a SUSY-conserving bilinear term is forbidden in the superpotential of the 
visible sector, a naturally small
$\mu$ proportional to the gravitino mass 
can be generated through interactions with hidden-sector fields.

In brief, the $\hat{C}_i$, $\hat{C}_i^{\dag}$, $\hat{h}$, and $\hat{h}^{\dag}$ are related by the minimal K\"{a}hler metric,
\begin{equation}
K(\hat{h},\hat{h}^{\dag},\hat{C}_i,\hat{C}_i^{\dag})=\hat{h}^{\dag}\hat{h}+\sum_i \hat{C}_i^{\dag}\hat{C}_i+\frac{\lambda}{\mpl}\hat{h}^{\dag}\hat{H}_u\hat{H}_d\,,
\label{Kahler}
\end{equation}
where $\mpl\simeq2\times10^{18}\gev$ is the reduced Plank mass and $\lambda$ is an arbitrary
(for our purpose) coupling constant. 

$N=1$ supergravity is spontaneously broken when the scalar components of the hidden sector
superfields develop vevs, $\langle h\rangle= M_X\sim\mpl$, 
while their $F$-components obtain vevs
$F\equiv\langle f_h\rangle\sim M_s^2 \mpl$, where $M_s\ll\mpl$ is the SUSY-breaking scale in the hidden sector, which we leave undetermined over here, and $f_h$
is defined up to an additive constant of the same order of $M_s^2 \mpl$. 
Since the K\"{a}hler metric~(\ref{Kahler}) 
is flat in the visible-sector fields, in the ``flat-space" limit, $\mpl\rightarrow\infty$, the 
scalar fields obtain universal masses $\mzero= m_{3/2}$,
where  
\be
m_{3/2}=e^{a^2/2}\frac{F}{\mpl^2}\label{gravitino}
\ee
is the gravitino mass and $a=M_X/\mpl$\,.

If the $\mu$ term is forbidden in the visible-sector superpotential by some symmetry,
as is the case, e.g., in the Missing Partner mechanism\cite{Masiero:1982fe,Grinstein:1982um} that we come back to in \refsec{sec:model},
the coupling with the hidden-sector field in Eq.~(\ref{Kahler}) generates an ``effective" $\mu$,
$\mu_{\textrm{eff}}$, in the superpotential, related to the gravitino mass\cite{Giudice:1988yz},
\be
\mu_{\textrm{eff}}= \frac{\lambda\mpl M_s^2}{F}m_{3/2}\,.
\label{muGM}
\ee

Since the same mechanism also generates masses for the scalar fields, $\mzero= m_{3/2}$,
$\mu_0$ ($\equiv\mu_{\textrm{eff}}$) becomes dynamically related to \mzero,
\be 
\mu_0=C_h m_0\,,\label{mugut}
\ee
where $C_h$ is a constant determined by the physics of the hidden sector.

Equation~(\ref{mugut}) can be used together with Eq.~(\ref{muscale}) to relate 
the value of the parameter $\mu$ at \msusy\ with the common scalar mass,
\be 
\mu=(R C_h) m_0=c_H \mzero\,,\label{mudown}
\ee
where now the effects of the hidden-sector physics are bundled with the running in the 
rescaled constant $c_H$.

One can quantify to what extent a relation like (\ref{mudown}) can reduce the fine tuning in the $\sim1\tev$
higgsino region. By using Eqs.~(\ref{expans}), (\ref{Defbf}), and (\ref{mudown}), one can recast Eq.~(\ref{natur}),
\begin{eqnarray} 
\frac{\partial\ln M_Z^2}{\partial\ln m_0^2}&\approx &2\frac{m_0^2}{M_Z^2}\left\{-\frac{\partial\mu^2}{\partial m_0^2}
-\frac{\partial\mhusq(\msusy)}{\partial m_0^2}\left[1+\mathcal{O}(10^{-2})\right]\right\}\nonumber\\
 &\approx &2\frac{m_0^2}{M_Z^2}\left(-c_H^2-0.64\,b_F^2 + 0.57\right)\,.\label{naturd}
\end{eqnarray}
In the $\mu\simeq 1\tev$ higgsino region \mzero\ and $c_H$ are related by Eq.~(\ref{mudown}).
By adjusting $b_F$ and $c_H$, one can in principle always 
find a region of parameter space for which the fine tuning induced by the value of $m_{3/2}$ achieves any desired size,
even for large \mzero.

\begin{figure}[t]
\centering
\subfloat[]{
\label{fig:a}
\includegraphics[width=0.50\textwidth]{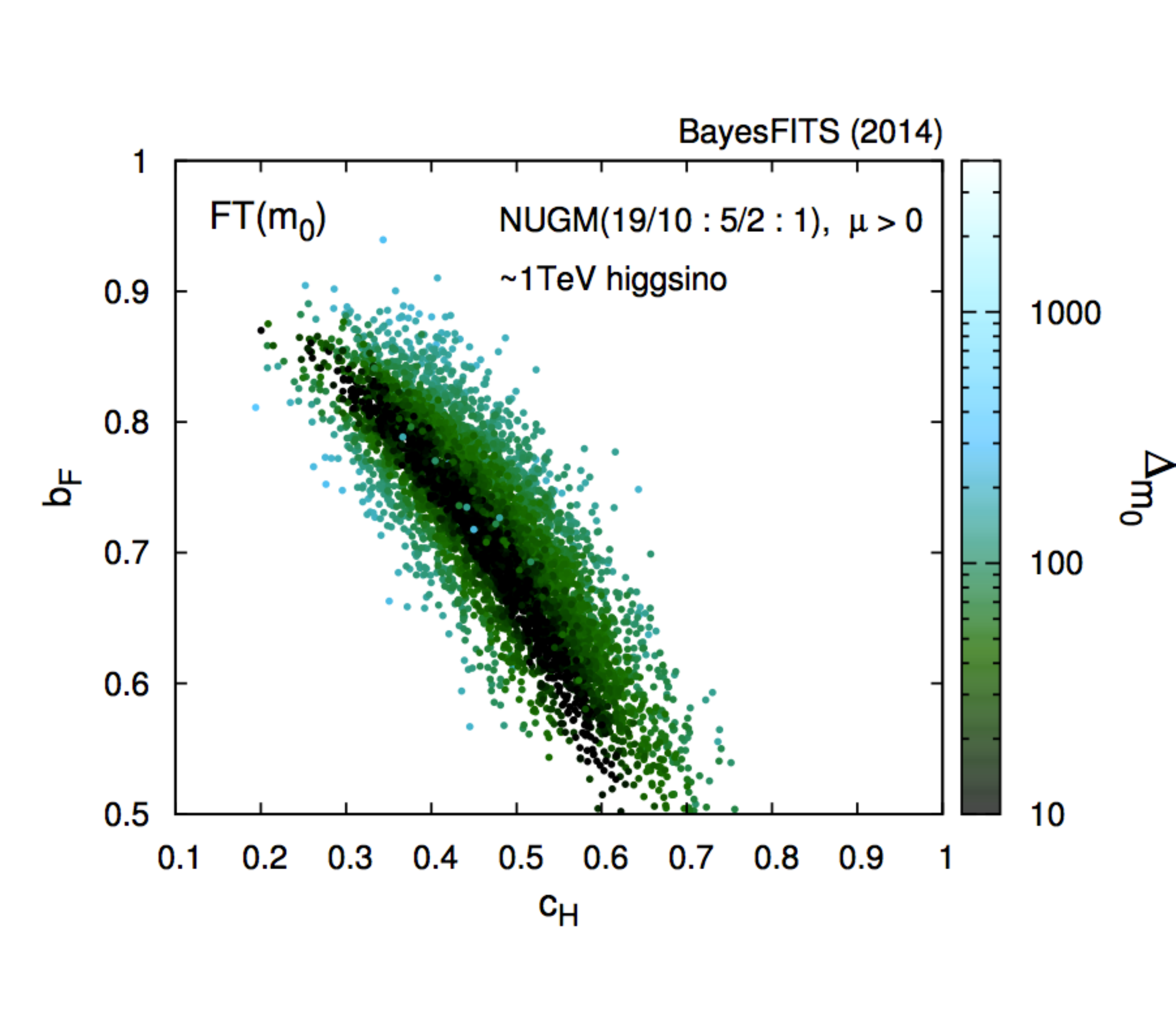}
}
\subfloat[]{
\label{fig:b}
\includegraphics[width=0.50\textwidth]{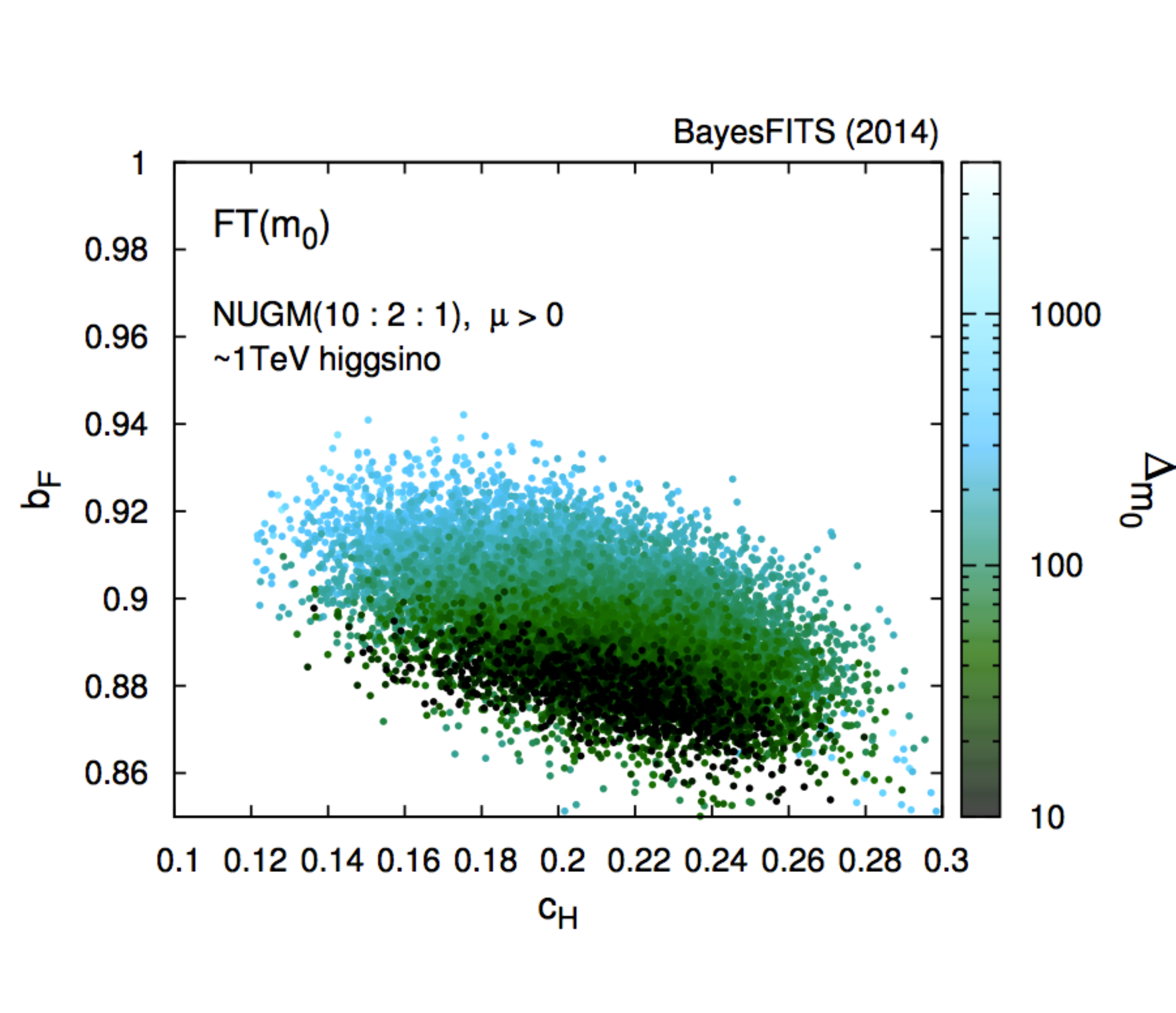}
}\\
\subfloat[]{
\label{fig:c}
\includegraphics[width=0.50\textwidth]{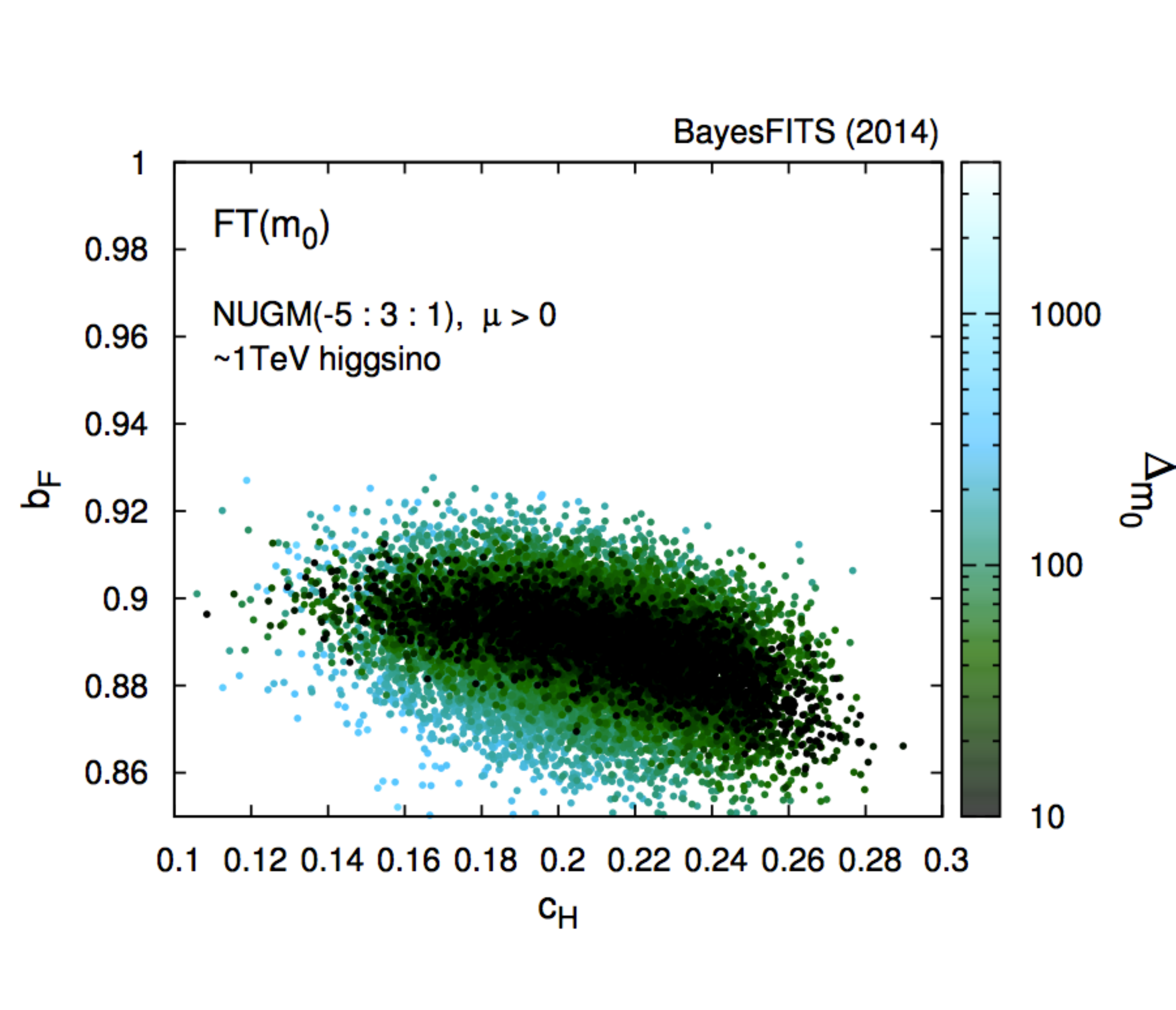}
}
\caption[]{\footnotesize Scatter plots of the scalar/$\mu$ fine tuning in the ($c_H$, $b_F$) plane 
for different gaugino mass patterns: \subref{fig:a} ($19/10:5/2:1$), \subref{fig:b} ($10:2:1$), and \subref{fig:c} ($-5:3:1$).
All points satisfy the constraints of Table~\ref{tab:exp_constraints} at $2\sigma$.}
\label{fig:Masiero}
\end{figure}

In \reffig{fig:Masiero} we show the distribution of $\Delta_{\mzero}$, Eq.~(\ref{naturd}),
in the ($c_H$, $b_F$) plane for the three NUGM choices that lead 
to the lowest gaugino fine tuning: \subref{fig:a} ($19/10:5/2:1$), \subref{fig:b} ($10:2:1$), and \subref{fig:c} ($-5:3:1$).
All of the points belong to the respective $\sim 1\tev$ higgsino regions
and satisfy the constraints of Table~\ref{tab:exp_constraints} at $2\sigma$.
In all the cases, the dark crescent-shaped areas with $\Delta_{\mzero}\ll 10$ show 
the values that a ($c_H$, $b_F$) pair can adopt to substantially reduce the fine tuning.
The difference between the three plots is almost entirely due to the phenomenology: as \reffig{fig:NUGM} shows,
the position of the $\sim1\tev$ higgsino region in 
the (\mzero, $M_3$) plane strongly depends on the chosen gaugino mass pattern, and this leads 
to differences in the allowed $c_H$ ranges. However, it is clear that $c_H$ and $b_F$ are related to each other
in a way largely independent of the gaugino mass pattern.

It is quite a striking coincidence, on the other hand, that 
the majority of points presenting low fine tuning in Figs.~\ref{fig:Masiero}\subref{fig:a} and 
\ref{fig:Masiero}\subref{fig:c}, and a significant fraction of the ones with low fine tuning in 
\reffig{fig:Masiero}\subref{fig:b}, lie right inside the 
region allowed by the phenomenological constraints, which features the correct Higgs mass and 1\tev\ higgsino dark matter.

Note also that it is necessary to have $b_F<1$ to obtain low levels of \mzero\ fine tuning, similarly to the cases where $\mu_0$ is a fundamental parameter, discussed in \refsec{sec:NUscan}.
The desired GUT-scale splitting between \mhu\ and the remaining scalar masses could be obtained, for instance, if one
considers a non-flat K\"{a}hler metric in the visible sector, instead of Eq.~(\ref{Kahler}).
Alternatively, the splitting could be generated through quantum gravitational corrections above the GUT scale\cite{Choi:1997de}, 
although it is far from clear to what extent such a procedure could by carried out in a well-defined and controllable way. 

If one neglects the gravitational quantum corrections, one way to control the amount of \mhu/\mzero\ splitting is to employ RGE running above the GUT scale, 
if a larger GUT gauge group breaks down to the SM group at \mgut.
Additional symmetries can be then assumed to forbid the $\mu$ term in the visible sector.
A possible solution of this kind in the context of $SU(5)$ is the subject of  
\refsec{sec:model}.

\begin{figure}[t]
\centering
\subfloat[]{
\label{fig:a}
\includegraphics[width=0.50\textwidth]{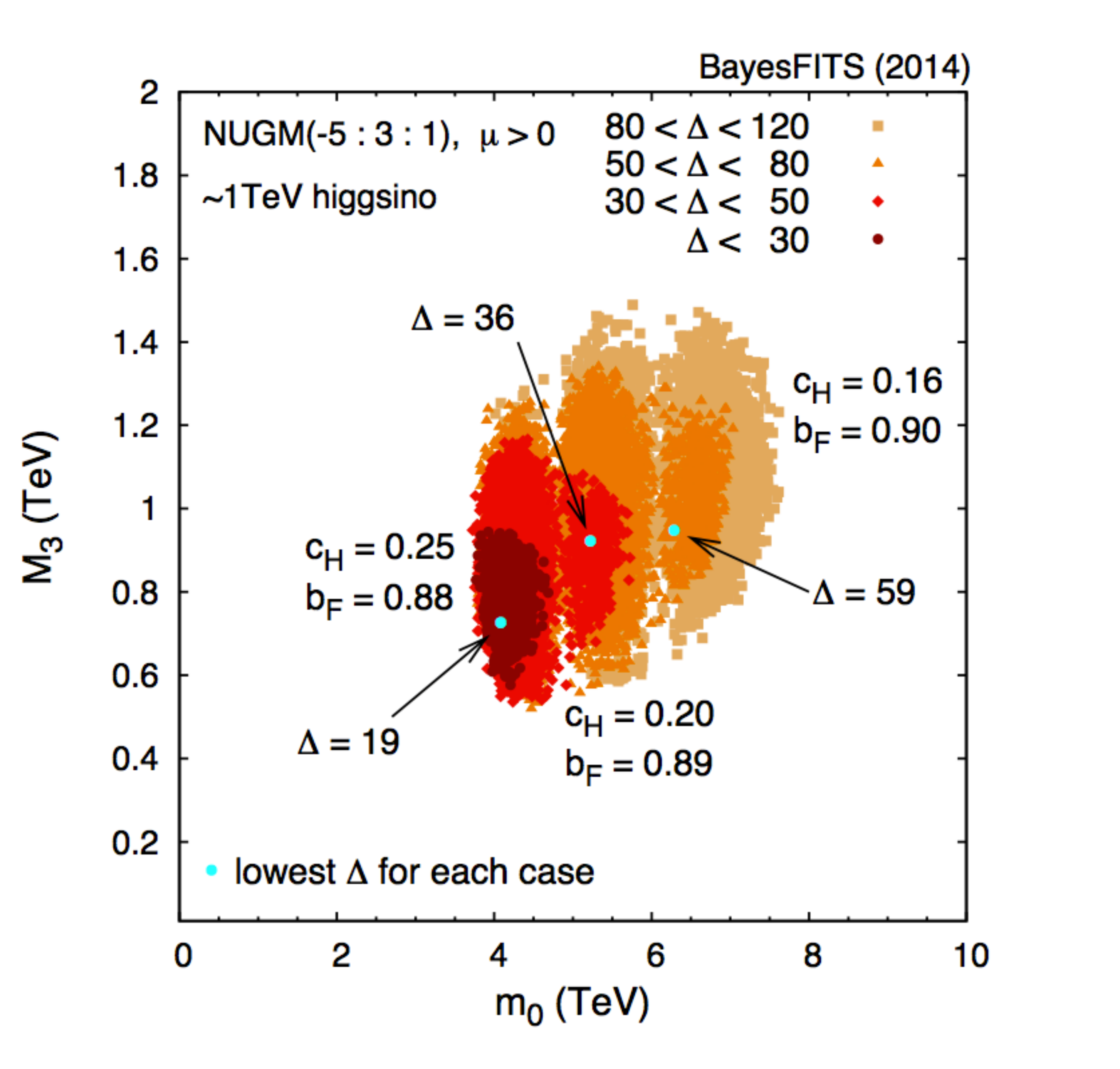}
}
\subfloat[]{
\label{fig:b}
\includegraphics[width=0.52\textwidth]{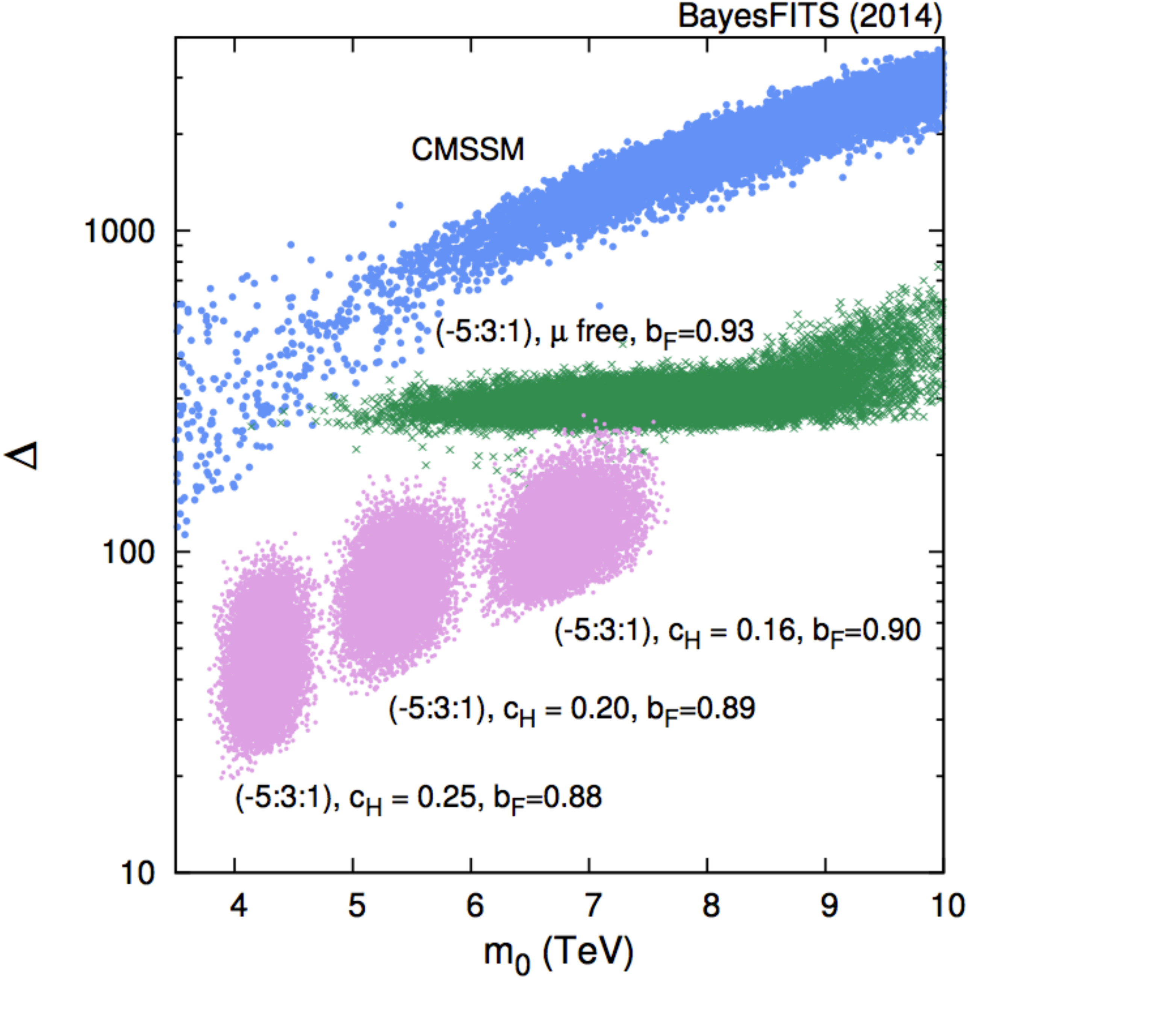}
}
\caption[]{\footnotesize \subref{fig:a} Regions of low fine tuning in the (\mzero, $M_3$) plane for different choices
of ($c_H$, $b_F$) in the NUGM ($-5:3:1$) case. \subref{fig:b} Fine tuning of the three models shown in \subref{fig:a} (small violet dots)
compared to (green crosses) the case shown in \reffig{fig:NUGM2}\subref{fig:b} ($\mu$ and \mzero\ unrelated) and (blue dots) the CMSSM.}
\label{fig:3regs}
\end{figure}

Finally, we present in \reffig{fig:3regs}\subref{fig:a} the distribution of the total fine tuning in the (\mzero, $M_3$)
plane for three different choices of ($c_H$, $b_F$) in the NUGM ($-5:3:1$) case.
One can see that values of $\Delta$ well below 100 can be obtained in all the cases. 
The lowest fine tuning values found in our scans were: 
$\Delta=19$ for $c_H=0.25, b_F=0.88$; $\Delta=36$ for $c_H=0.20, b_F=0.89$; and $\Delta=59$ for $c_H=0.16, b_F=0.90$.
Note that in all the considered cases the dominant contribution to $\Delta$ is given by $\Delta_{M_3}$,
so that indeed Eq.~(\ref{mugut}) can reduce the fine tuning due to \mzero\ very efficiently. 

In \reffig{fig:3regs}\subref{fig:b} we compare the fine tuning of these three cases (small violet dots) with that of the NUGM case first shown in 
\reffig{fig:NUGM2}\subref{fig:b} (green crosses), which was featuring the same gaugino mass pattern, 
$b_F=0.93$ to cancel the scalar-sector fine tuning, and 
$\mu_0$ as an independent parameter of the theory. We also show the fine tuning of the CMSSM
(blue dots) for comparison. 
The $\Delta$ of the case with $b_F=0.93$ is dominated by $\Delta_{\mu}\simeq 250$, typical of the
$\sim 1\tev$ higgsino region. 
One can see that relating $\mu_0$ to \mzero\ reduces the fine tuning by approximately
one order of magnitude if $c_H$ is chosen so as to give a $\sim 1\tev$~higgsino region at $\mzero\simeq 4\tev$.
For different $c_H$ values, which would lead to higgsino regions with larger \mzero, the fine tuning can still be 5 times smaller than
$\Delta\simeq 250$. 
In contrast, the fine tuning of the $\sim 1\tev$ higgsino region can be reduced, for the largest \mzero\ values shown here,
by 15--20 times relative to the CMSSM.

\subsection{Spectra and phenomenology}\label{sec:pheno}

\begin{figure}[t]
\centering
\subfloat[]{
\label{fig:a}
\includegraphics[width=0.30\textwidth]{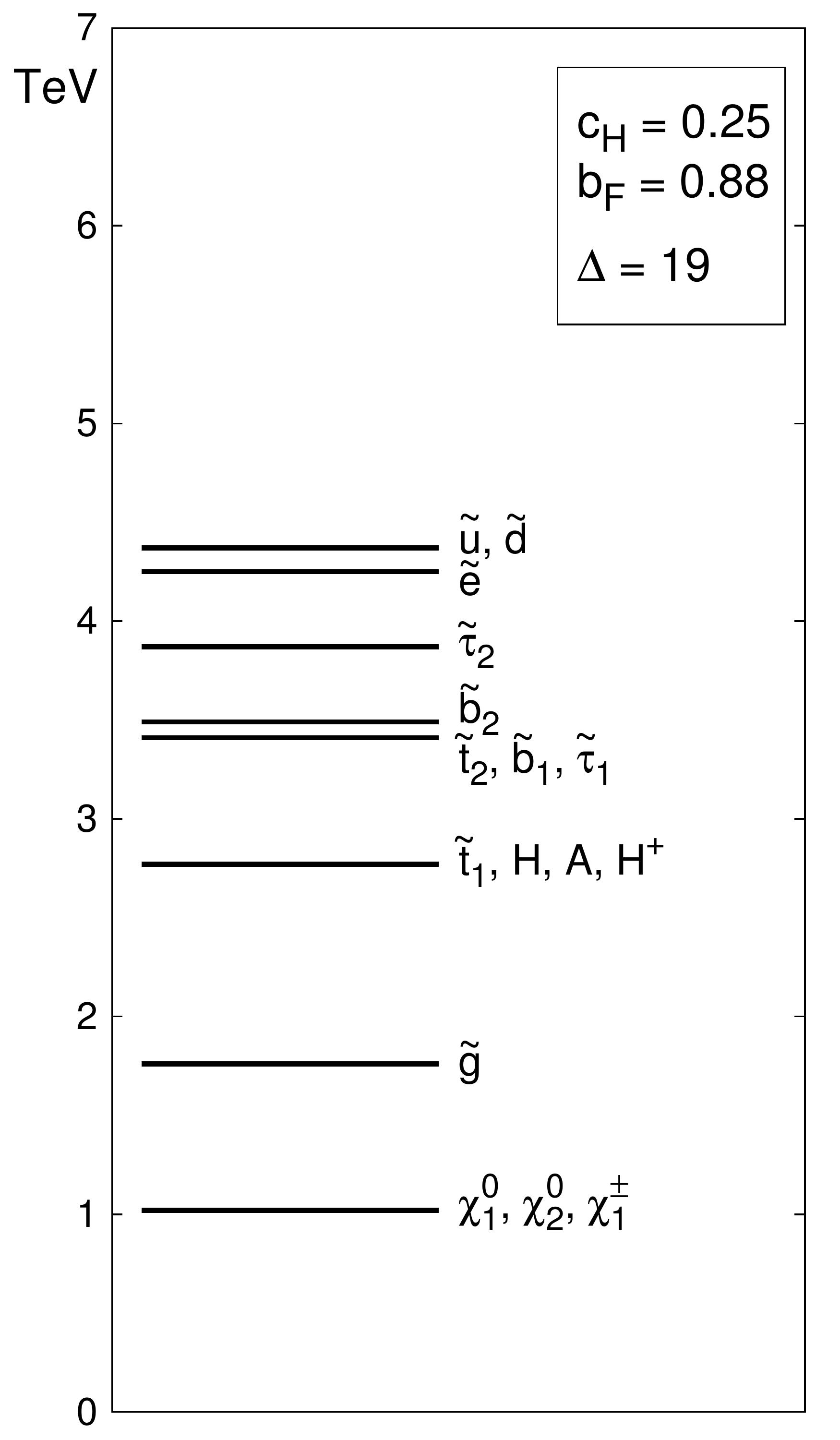}
}
\subfloat[]{
\label{fig:b}
\includegraphics[width=0.30\textwidth]{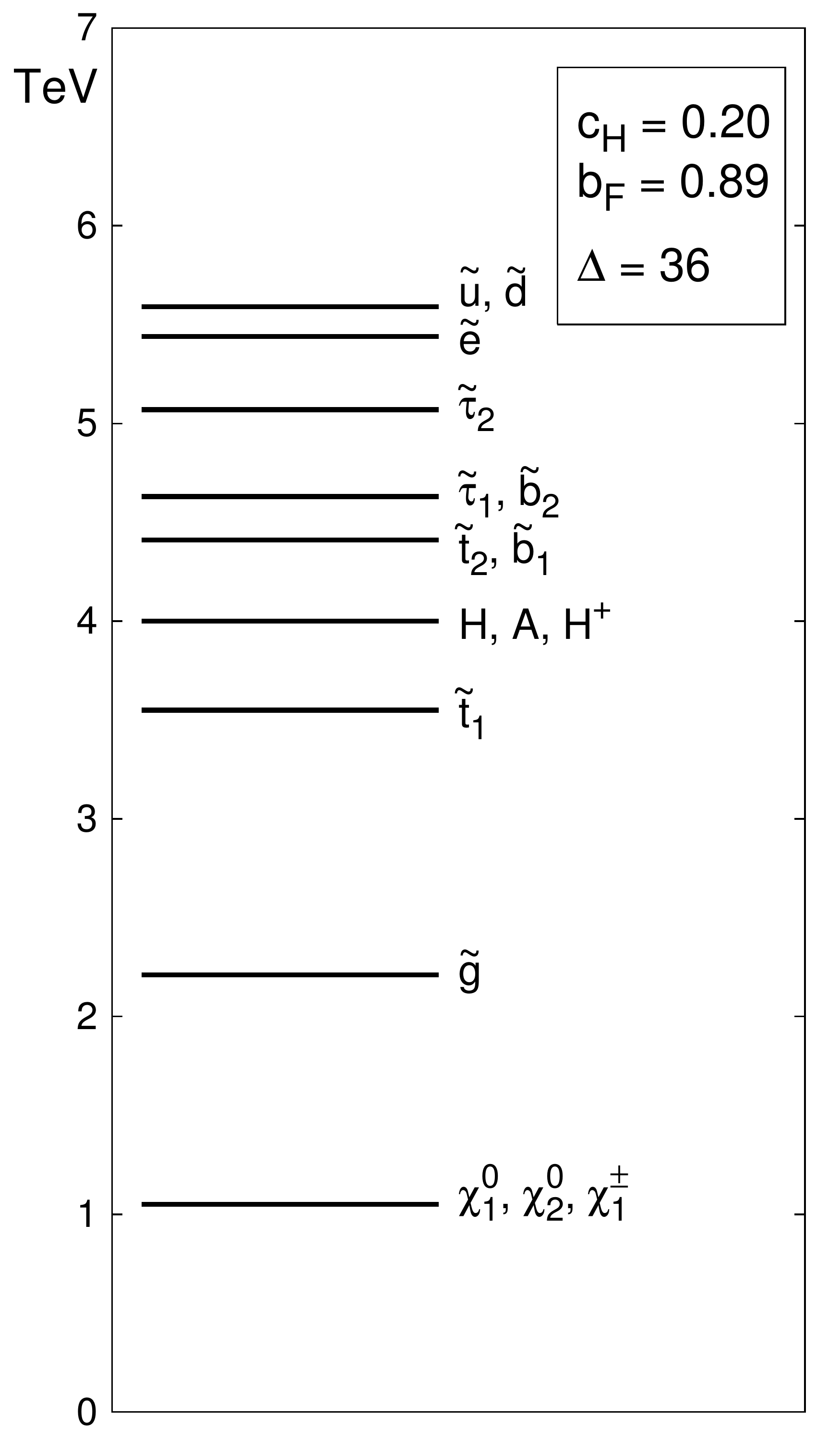}
}
\subfloat[]{
\label{fig:c}
\includegraphics[width=0.30\textwidth]{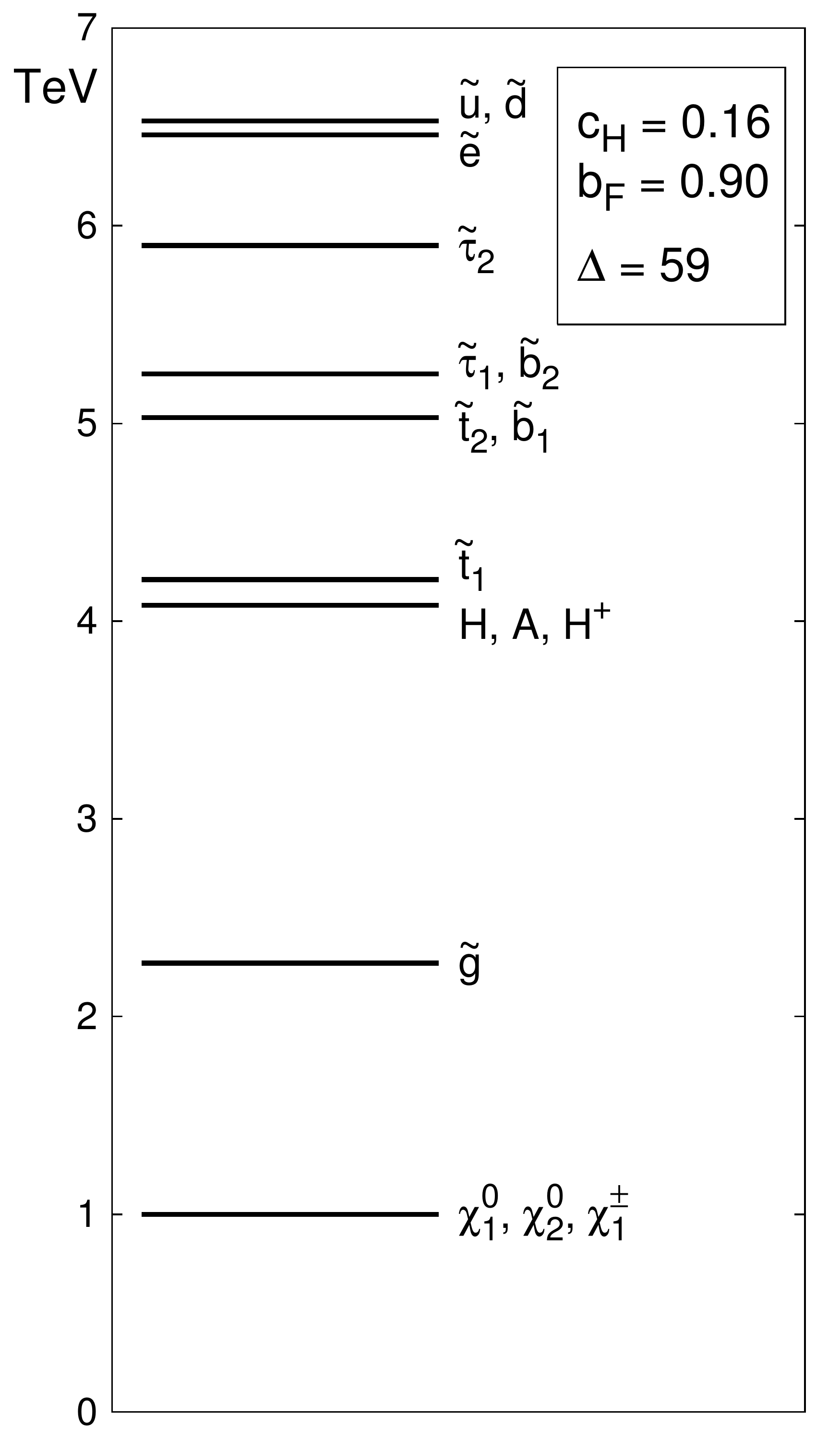}
}
\caption[]{\footnotesize Supersymmetric mass spectra for the points of lowest $\Delta$ 
shown in \reffig{fig:3regs}\subref{fig:a}.}
\label{fig:spectra1}
\end{figure}

In \reffig{fig:spectra1}\subref{fig:a} we show the spectrum of the point with lowest 
$\Delta$ for $c_H=0.25$, $b_F=0.88$ in the NUGM ($-5:3:1$). 
The spectra for $c_H=0.20$, $b_F=0.89$  and $c_H=0.16$, $b_F=0.90$ are shown
in \reffig{fig:spectra1}\subref{fig:b} and \reffig{fig:spectra1}\subref{fig:c}, respectively.

Obviously, the scenario that shows the better prospects is the one characterized 
by lighter sparticles, shown in \reffig{fig:spectra1}\subref{fig:a}. Even in that case, though,
the requirement of good relic density narrows down the neutralino mass to
$\mchi\simeq 1\TeV$, a value that will provide a challenge for observation of other superpartners
at the LHC, as it strongly limits the 
transverse momentum of the charged and colored SUSY particles produced in collisions. 

From this perspective, it does not seem surprising that SUSY particles have not been observed so far at the LHC and we fear that,
if naturalness happened to be encoded in SUSY the way we analyzed in this paper, there will probably be little chance to see 
sparticles even in future runs.

Rather than at the LHC, the best prospects for observation of this kind of scenarios come from dark matter direct detection
experiments, particularly at 1-tonne detectors like XENON1T\cite{Akerib:2012ys}. 
It has been shown, see e.g.,\cite{Fowlie:2013oua}, that there are good prospects for future detection of an $\mchi\simeq1\tev$ neutralino. 
We present in Table~\ref{tab:DDvalues} the values of the spin-independent neutralino-proton 
cross section for the points of lowest $\Delta$ in the three cases given above.

Unfortunately, since these scenarios have approximately all the same \mchi\ and the same higgsino composition, even 
upon detection at 1-tonne detectors it will be hard to distinguish one case from another, so that the values of the scalar 
masses will have to be inferred by other means. However, even if the values shown in Table~\ref{tab:DDvalues} are subject to some theoretical
uncertainties, they seem to indicate that an eventual detection at XENON1T is more likely for the cases of larger $c_H$ (smaller \mzero).

\section{Low fine tuning from high-scale effects}\label{sec:model}

We have shown in \refsec{sec:muterm} that a simple relation between $\mu_0$ and \mzero\ can increase the naturalness
of the $\sim 1\tev$~higgsino region of the MSSM.
We assumed that the origin of Eq.~(\ref{mugut}) lay in some high-scale physics or hidden-sector mechanism,
and we referred to Giudice-Masiero as the most widely known example.
However, a relation like (\ref{mugut}) will not make the model more natural
without additional conditions. As \reffig{fig:Masiero} shows, $b_F$ also must assume a value that conspires 
with the particular choice of $c_H$, and this value certainly cannot be equal 1.
Finally, there is the fine tuning of the gaugino sector that can be kept under control if one assumes
certain non-universality condition for the gaugino masses. 
We have shown in \refsec{sec:NUGM} that, in the context of GUT unification, 
some patterns emerging from $SU(5)$ are among the most efficient in this sense. 

In this section we address the issue of obtaining an appropriate value of $b_F$
at the GUT scale in the context of $SU(5)$, maintaining consistency with the Giudice-Masiero mechanism. 
We consider a minimal supergravity scenario in which supersymmetry breaking is transmitted
to the visible sector at the generic scale $M_X$. $M_X$ can be identified with the GUT scale, $\mgut\simeq 2\times10^{16}\gev$ (as
it happens in the CMSSM), but also with any scale above, up to the reduced Plank scale,
$\mpl\simeq 2\times10^{18}\gev$.
We assume that between \mgut\ and $M_X$ the theory has an extended gauge symmetry, 
which we choose to be $SU(5)$ for consistency with requirements of low gaugino fine tuning in the MSSM.  
Thus, the high-scale parameters are renormalized when evolved down to the GUT scale, 
leading to non-universal scalar masses at \mgut,
which can in turn lead to the appropriate low fine tuning if $b_F$ assumes the correct value.
 
\begin{table}[t]\footnotesize
\begin{center}
\begin{tabular}{|l|l|l|l|}
\hline
($c_H$, $b_F$) & $\Delta$ & \mchi\ (GeV) & \sigsip\ (pb)\\
\hline
(0.25, 0.88) & 19 & 1021 &  $6.22\times10^{-10}$ \\
\hline
(0.20, 0.89) & 36 & 1048 &  $2.11\times10^{-10}$ \\
\hline
(0.16, 0.90) & 59 & 1005 &  $1.67\times10^{-10}$ \\
\hline
\end{tabular}
\caption{\footnotesize
The values of the spin-independent neutralino-proton cross section, \sigsip, for the points shown in 
\reffig{fig:spectra1}. For $\mchi\simeq1\tev$, the sensitivity of LUX is estimated
to reach $\sim2\times10^{-9}$~pb by the end of 2014\cite{Akerib:2012ys} and the one of XENON1T 
should reach $\sim2\times10^{-10}$~pb by the end of 2017\cite{Aprile:2012zx}.
} 
\label{tab:DDvalues}
\end{center}
\end{table}
 
We additionally assume that the doublet-triplet splitting problem of $SU(5)$ is solved by the Missing Partner mechanism\cite{Masiero:1982fe,Grinstein:1982um},
which can be used here to forbid the $\mu$ term in the superpotential of the visible sector superfields. 
An anomalous $U(1)_A$ symmetry  with a 
nonvanishing trace over the charges of the matter superfields is required to assure that the EW Higgs doublets are massless. 
Such a symmetry can be either global\cite{Altarelli:2000fu} or local\cite{Berezhiani:1996nu}. 
In the latter case, a non-zero Fayet-Iliopoulos term $\xi\propto \textrm{Tr}\,Q\,M_X^2$ is generated\cite{Witten:1981nf,Fischler:1981zk,Dine:1987xk,Atick:1987gy,Dine:1987gj} 
and some of the scalar fields can obtain a non-zero vev of order $\sim\sqrt{\xi}$\,.

We refer here to the model introduced in\cite{Berezhiani:1996nu}, but alternative choices could be employed to make our point, as long
as they are characterized by two important ingredients: \textit{a}) an extended gauge group compatible with one of the 
gaugino mass patterns that lead to low fine tuning and \textit{b}) a symmetry to forbid the $\mu$ term in the superpotential.
One could consider, for example, the use of a discrete $R$-symmetry instead of an anomalous $U(1)$,
but a series of no-go theorems\cite{Fallbacher:2011xg} prevent this choice in a simple 4-dimensional setup.

The superfield contents of the theory is the following\cite{Masiero:1982fe}: 
MSSM matter superfields are indicated with \ptenhat\ and \ffivehat\ as they belong to the \textbf{10} and $\bar{\mathbf{5}}$ representation,
respectively. Gauginos and gauge bosons belong to the {\bf 24} adjoint (and we indicate them by $\hat{\lam}$), 
Higgs doublets belong to the {\bf 5} and $\bar{\mathbf{5}}$ (\hfivehat, \hbfivehat) and they are accompanied by two color triplets. 
There are also additional Higgs multiplets in the {\bf 50} (\thethat) and the $\overline{\mathbf{50}}$ (\thetbhat), 
as well as one in the {\bf 75} ($\hat{\si}$) that breaks $SU(5)$ at the GUT scale. 
Finally, there is a gauge singlet $\hat{S}$ that breaks $U(1)_A$ at a scale $\widetilde{M}$. 

The most general renormalizable superpotential invariant under $SU(5)\times U(1)_A$ takes the form:
\begin{flalign}\label{superp}
W_{SU(5)\times U(1)}&=\mu_{\si}\textrm{tr}\hat{\si}^2+\frac{1}{6}\lam_{\si}\textrm{tr}\hat{\si}^3+\lams \hat{S}\thethat\thetbhat+\lamh\hfivehat\hat{\si}\thethat+\lamhb\hbfivehat\hat{\si}\thetbhat\nonumber\\
&+\frac{1}{4}Y_U\epsilon_{ijklm}\ptenhat^{ij}\ptenhat^{kl}\hfivehat^m+\sqrt{2}Y_D\ptenhat^{ij}\hat{\phi}_{5i}\hat{h}_{\bar{5}j},
\end{flalign}
with standard meaning of the symbols. 
The coupling constants are of order 1 and the mass term $\mu_{\si}$ is of the order of \mgut, the scale where $SU(5)$ is spontaneously broken.

The $U(1)_A$ charges can be chosen to forbid the bilinear term $\mu_H\hfivehat\hbfivehat$\cite{Altarelli:2000fu,Berezhiani:1996nu}:
\be
q_{\pten}=\frac{1}{2},\quad q_{\ffive}=\frac{1}{2},\quad q_{\hfive}=-1,\quad q_{\hbfive}=-1,\quad q_{\thet}=1, \quad q_{\thetb}=1, \quad q_{\si}=0, \quad q_{S}=-2\,. 
\ee
In fact, in order to have a relation between $\mu_0$ and \mzero,
we assume that the $\mu$ term of the superpotential is effectively generated by interaction with the hidden sector via the Giudice-Masiero mechanism. 
Note that not only the term $\mu_H\hfivehat\hbfivehat$, but also $\hat{S}\hfivehat\hbfivehat$ is forbidden in Eq.~(\ref{superp}) by the $U(1)_A$ symmetry.  
It has been shown\cite{Berezhiani:1996nu} that this solution is protected against Planck scale corrections to all orders, 
since non-renormalizable terms of the form $\sim 1/(\mpl^{k+n-1})\,\hat{S}^k\hat{\si}^n\hfivehat\hbfivehat$ are not allowed.

Finally, the soft SUSY-breaking part of the Lagrangian reads:
\begin{flalign}
\mathcal{L}_{\textrm{soft}}&=\mfive|\ffive|^2+\mten\textrm{tr}[\pten^{\dagger}\pten]+\dfive|\hfive|^2+\dbfive|\hbfive|^2\nonumber\\
&+\dfifty|\thet|^2+\dfiftyb|\thetb|^2+\msi\textrm{tr}[\si^{\dagger}\si]+m_S^2 |S|^2\nonumber\\
&+\frac{1}{6}A_{\si}\lam_{\si}\textrm{tr}\si^3+A_S\lams S\thet\thetb+ A_{H}\lamh\hfive\si\thet+A_{\bar{H}}\lamhb\hbfive\si\thetb \nonumber\\
&+\frac{1}{4}A_UY_U\epsilon_{ijklm}\pten^{ij}\pten^{kl}\hfive^m+\sqrt{2}A_DY_D\pten^{ij}\phi_{5i}h_{j\bar{5}}
+B_{\si}\mu_{\si}\textrm{tr}\si^2+\frac{1}{2}M_5\lam_a\lam_a+\frac{1}{2}M_{1_A}|\lam_{1_A}|^2\,.
\end{flalign}

The $U(1)_A$ symmetry is spontaneously broken when the singlet $S$ develops a vev and the Higgs multiplets 
\thet\ and \thetb\ receive a mass, $\widetilde{M}=\lams\langle S\rangle$. 
The $SU(5)$ symmetry is broken down to the MSSM $SU(3)\times SU(2)_L\times U(1)_Y$ when the field $\si$ 
develops a vev, $v_{\si}\simeq\mgut$. 
As a consequence, the Higgs fields and gauge bosons associated with the broken generators acquire masses of the order of $v_{\si}$. 
The masses of the color triplets of \hfive\ and \hbfive\ are generated via a ``seesaw" kind of mechanism,
since they mix with the corresponding triplets of \thet. The resulting effective triplet mass is 
\be
M_{H_3}=\frac{\lamh\lamhb v_\si^2}{\widetilde{M}}\,.
\label{h3mass}
\ee

The mass $M_{H_3}$ must satisfy the limits from proton stability, 
obtained by suppressing the dimension-five operators generated by the color
triplet\cite{Dimopoulos:1981dw,Ellis:1981tv,Sakai:1981pk,Weinberg:1981wj,Hisano:1992jj}. 
Given the present experimental 90\%~C.L. lower bound on the proton partial lifetime in the decay mode $p\to K^+\bar{\nu}$,  
$\tau(p\to K^+\bar{\nu})>2.3\times 10^{33}$ yrs\cite{Beringer:1900zz}, it can be derived\cite{Hisano:1992jj} that 
\be\label{m3lim}
M_{H_3}>2.4\times 10^{16}\gev\,.
\ee

Recalling that $v_\si\simeq\mgut$, Eq.~(\ref{m3lim}) effectively constrains the mass $\widetilde{M}$ to
\be\label{mtillim}
\mgut\leq\widetilde{M}\leq(\lamh\lamhb)\frac{\mgut^2}{2.4\times 10^{16}\gev}\,,
\ee
which translates into a lower bound for \lamh\lamhb: 
\be\label{lamlim}
\lamh\lamhb\geq\frac{2.4\times 10^{16}\gev}{\mgut}\simeq 1.2\,.
\ee

The doublet components of \hfive\ and \hbfive\ do not mix with any other field and they thus remain massless, unless the Giudice-Masiero mechanism
is used to generate the mass term for the EW Higgs doublets.   
The choice of K\"{a}hler metric given in Eq.~(\ref{Kahler}) leads to the universal boundary conditions in the scalar sector at the scale $M_X$,
\begin{equation}
m_0^2\equiv\mfive=\mten=\dfive=\dbfive=\dfifty=\dfiftyb=\msi=m_S^2\,.
\end{equation}
The RGEs for the couplings and soft-SUSY breaking parameters are given in Eqs.~(\ref{rgesu5}) of Appendix~\ref{app:a}. 
One can use Eqs.~(\ref{rgesu5}) to derive approximate expressions for the following soft masses at the scale \mgut\,:
\begin{eqnarray}
\mfive(\mgut)&\approx &\left[\mzero^2-\frac{48}{30}\left(1-\frac{1}{(1+3g_5^2t/8\pi^2)^2}\right)M_5^2\right]\exp\left(\frac{3}{2\pi^2}Y_D^2t\right)\,,\nonumber\\
\mten(\mgut)&\approx &\left[\mzero^2-\frac{72}{30}\left(1-\frac{1}{(1+3g_5^2t/8\pi^2)^2}\right)M_5^2\right]\exp\left[\left(\frac{9}{8\pi^2}Y_U^2+\frac{3}{4\pi^2}Y_D^2\right)t\right]\,,\nonumber\\
\dfive(\mgut)&\approx &\left[\mzero^2-\frac{48}{30}\left(1-\frac{1}{(1+3g_5^2t/8\pi^2)^2}\right)M_5^2\right]\exp\left[\left(\frac{9}{8\pi^2}Y_U^2+\frac{45}{8\pi^2}
\lamh^2\right)t\right]\,,\nonumber\\
\dbfive(\mgut)&\approx &\left[\mzero^2-\frac{48}{30}\left(1-\frac{1}{(1+3g_5^2t/8\pi^2)^2}\right)M_5^2\right]\exp\left[\left(\frac{3}{2\pi^2}Y_D^2+\frac{45}{8\pi^2}
\lamhb^2\right)t\right]\,,
\end{eqnarray}
where $t=\ln(\mgut/M_X)$ and we have assumed that $g_{1_A}$ is small enough so that its contribution can be neglected in a first approximation. We also 
neglected the trilinear terms, which are relatively small with respect to 
\mzero, see relation (\ref{ranges2}).

The RGE-driven mass difference between the Higgs soft mass and the other scalars at the 
GUT scale is given by the ratio $\dfive(\mgut)/\mten(\mgut)$ and is entirely determined by four parameters: 
the superpotential trilinear Higgs and Yukawa couplings, $\lamh$ and $Y_D$, the scale difference, $t$,
and the ratio of the common gaugino to the common scalar mass, $M_5/\mzero$. 
Assuming that $Y_D\ll\lamh$ one can write:
\begin{eqnarray}\label{ratio}
\frac{\dfive(\mgut)}{\mten(\mgut)}&=&F(t,\mzero,M_5)\exp\left(\frac{45}{8\pi^2}\lamh^2 t\right)\,
\end{eqnarray}
and the function $F(t,\mzero,M_5)\approx 1$ in the limit where $\mzero>M_5$ 
and $|t|<2$. 

In the three cases shown in Fig.~\ref{fig:3regs}\subref{fig:a} the first condition is satisfied 
($\mzero>M_3$, $M_5\approx M_3$ after $SU(5)$ breaking). The second relation also holds, due to both the requirement of proton stability 
and the large Higgs representations present in the model. The other mass ratios are essentially stable under the running since the Yukawa 
couplings $Y_U$ and $Y_D$ are smaller than $\lamh$ and they couple to sfermions through the Higgs fields belonging to much smaller representations.

By using Eqs.~(\ref{h3mass})--(\ref{lamlim}), one can derive the implications of our chosen values of 
$b_F^2\equiv\dfive(\mgut)/\mten(\mgut)$ on the physics above \mgut.
We assume that $\lamh^2\simeq\lamh\lamhb\simeq 1.2$ so that, consequently, $\widetilde{M}\simeq\mgut$.
Note that this assumption makes the gauge coupling $g_5$ become non-perturbative well before reaching the Planck scale, at $M_{\textrm{pert}}=9\times10^{16}\gev$
(one can use the first of Eqs.~(\ref{rgesu5}) to prove this). From Eq.~(\ref{ratio}),
\be\label{mlrel}
\frac{M_X}{\mgut}=b_F^{-16\pi^2/(45\cdot 1.2)}\,.
\ee

For the three cases of Fig.~\ref{fig:3regs}\subref{fig:a}, Eq.~(\ref{mlrel}) gives:
$M_X\simeq1.36\,\mgut$ for $b_F=0.90$, $M_X\simeq1.40\,\mgut$ for $b_F=0.89$, and $M_X\simeq1.45\,\mgut$ for $b_F=0.88$. 
The scale of supersymmetry breaking appears to be quite close to the GUT scale. 
Incidentally, in supergravity the trilinear terms are approximately given by 
$|\azero|\sim a\,m_{3/2}$, with $a=M_X/\mpl$ like in Eq.~(\ref{gravitino}). Thus, the values of $M_X$ we obtained also 
help respect the \azero\ ranges chosen in our scans, see the first of (\ref{ranges2}).\footnote{The desired $B_0$ (\tanb) ranges are more difficult to enforce
in supergravity.
They can be obtained, nonetheless, possibly at the expense of introducing a large cosmological constant.}

The scale of supersymmetry breaking could be, in principle, moved up if one allows for $\lamh\ne\lamhb$. 
In particular, if $\lamhb=2.3$ (and $\lamh=0.53$) $M_X$ can be taken equal to the perturbativity scale, $M_{\textrm{pert}}$. 
Such a choice, however, results in a very strong renormalization of the mass \dbfive, as it is affected by \lamhb, 
leading to $\Delta_{\bar{5}}\simeq 0.1\,m_{10}$ at the GUT scale. The latter would imply $\mhd(\mgut)/\mzero\simeq 0.1$, 
a condition that often leads to no EWSB at the low scale.

Note, finally, that the mechanism described here can be used beyond the need of relating
$\mu_0$ and \mzero. As was shown in \refsec{sec:gut_fine}, even by accepting the presence of an independent $\mu$
parameter, $b_F\simeq 0.93$ is required to reduce the fine tuning of the scalar sector with respect to the
CMSSM. To generate the right $b_F$ in that case, however, it might be enough to use a model characterized by smaller representations of the gauge group,
as it happens, e.g., in minimal $SU(5)$\cite{Polonsky:1994rz}. 
In that case one could run the parameters further up above \mgut, 
to energies very close to \mpl. However, large trilinear terms would have then to be forbidden by other means, 
or their relation to the scalar masses by considered in the calculation of the overall fine tuning due to $m_{3/2}$\,.

\section{Summary and conclusions}\label{sec:con}


In this paper we investigated the issue of fine tuning in the MSSM with GUT-scale 
boundary conditions. We analyzed several popular cases: the CMSSM, models with non-universal gaugino masses, 
and models with non-universal scalar masses.
We employed the widely accepted Barbieri-Giudice measure, $\Delta$, to quantify the fine tuning, and we identified the GUT-scale relations among the input 
parameters that led to small $\Delta$.
We focused in particular in the region of the parameter space characterized by the correct Higgs mass, production and decay rates in good agreement with the 
experimental value, and for which the relic density constraint is satisfied thanks to a nearly pure higgsino LSP at
$\mchi\simeq 1\tev$. This region is also characterized by stop masses in the multi-TeV
regime, so that its fine tuning can be very large if one assumes the simplest unification conditions like in the CMSSM. 

We showed that the mechanism of parameter-focusing along RGE running is, for certain specific 
GUT-relations, very efficient in lowering the fine tuning of the scalar and gaugino sectors in the $\sim 1\tev$~higgsino region
with respect to the case with universal masses.
In particular, very low fine tuning can be obtained in the gaugino sector if one adopts the
$SU(5)$ condition $(M_1:M_2:M_3)=(-5:3:1)$. And in the scalar sector, the fine tuning is strongly reduced if the condition 
$\mhu(\mgut)/\mzero\simeq 0.92-0.94$ is satisfied at the GUT scale, even for regions 
where \msusy\ is in excess of $5-6\tev$.

However, even when the above patterns are enforced, there still remains the issue of a $\mu$ parameter too large to 
give an appreciably small $\Delta$, as $\mu\simeq 1\tev$ implies $\Delta$ of the order of a few hundreds. 
This may be one of the reasons why in the literature the $\sim 1\tev$ higgsino region has so far been neglected in studies of low fine tuning. 
We showed in this paper that the issue can be approached in a way no different from the other soft-breaking parameters 
if one assumes the existence of a relation 
among $\mu$ and some other soft masses at the GUT scale.
In particular we showed that, if $\mu$ and the GUT-scale scalar mass \mzero\ 
are linearly related, the $\mu$ fine tuning becomes greatly reduced with respect to the case where $\mu$ and the soft masses are unrelated.
We quantified the improvement in fine tuning and we finally provided one example of how 
a $\mu_0\sim\mzero$ relation can be generated, at least at the tree level:
by combining the known Masiero-Giudice mechanism in supergravity and the Missing Partner mechanism in 
$SU(5)$ GUT.

We point out here that the example we employ is not free of problems, as is often the case in supergravity where
the tree-level relations that lead to universality are spoiled by large and mostly uncontrollable radiative corrections.
Moreover, it would be interesting to investigate the consistency of our parameter ranges with the measured value of the cosmological constant
in a supergravity-inspired theory like this one, 
an issue that goes beyond the scope of this paper.
It is quite possible that the real high-scale theory will eventually turn out to be different from the simplified case 
discussed here, it might be perhaps based on some string theory.

More importantly, one can argue that by introducing the parameters $c_H$ and $b_F$ the
problem of fine tuning in the MSSM is not solved, but rather shifted to the framework of a higher scale theory. 
Particularly, even if the EW scale
remains stable under the changes of the GUT-defined MSSM soft terms, which is in itself an achievement worth the effort,
the fine tuning will most likely be reintroduced under variations of the continuous fundamental parameters of the underlying theory.
The problem can be avoided if a symmetry or some other dynamical mechanism is responsible 
for the constants that relate the scalar and $\mu$ sectors. 
Thus the effort in this direction should continue.

What we gave here is a set of guidelines to the realizations
of high-scale physics that are allowed if the effective
low-scale theory is not supposed to be excessively fine-tuned.
As was pointed out in \refsec{sec:GMmech}, it is striking that the values of $c_H$ and $b_F$ that 
emerge from requiring a higher level of naturalness are in many occasions also the ones that are 
favored by independent phenomenological constraints.

\begin{center}
\textbf{ACKNOWLEDGMENTS}
\end{center}

  We would like to thank Gian F. Giudice and Graham G. Ross for helpful comments on the manuscript. 
  This work has been funded in part by the Welcome Programme
  of the Foundation for Polish Science.
  K.K. is supported by the EU and MSHE Grant No. POIG.02.03.00-00-013/09.
  L.R. is also supported in part by a STFC
  consortium grant of Lancaster, Manchester, and Sheffield Universities. The use
  of the CIS computer cluster at the National Centre for Nuclear Research is gratefully acknowledged.
\newpage
\appendix
\section{One-loop RGEs for $\mathbf{SU(5)\times U(1)}$}\label{app:a}
We present in this appendix the one-loop RGEs for the 
superpotential and soft SUSY-breaking parameters of the $SU(5)\times U(1)$ gauge group considered in \refsec{sec:model}.
We limit ourselves to the third generation for the Yukawa and $A$ terms. 
The results were obtained using the formulas of\cite{Martin:1993zk} for the $\beta$ functions for a generic gauge group 
and the numerical package \texttt{Susyno}\cite{Fonseca:2011sy}. 
In all equations we define $t=\ln(Q/\mgut)$.

\begin{flalign}
&\frac{d}{dt}g_5=\frac{1}{16\pi^2}52g_5^3,\nonumber\\
&\frac{d}{dt}g_{1_A}=\frac{1}{16\pi^2}\frac{501}{4}g_{1_A}^3,\nonumber\\
&\frac{d}{dt}M_5=\frac{1}{16\pi^2}104g_5^2M_5,\nonumber\\
&\frac{d}{dt}M_{1_A}=\frac{1}{16\pi^2}\frac{501}{2}g_{1_A}^2M_{1_A},\nonumber\\
&\frac{d}{dt}\mfive=\frac{1}{16\pi^2}[8Y_D^2(A_D^2+\dbfive+\mten+\mfive)-\frac{96}{5}g_5^2M_5^2-2g_{1_A}^2M_{1_A}^2+g_{1_A}^2m^2_{Y}],\nonumber\\
&\frac{d}{dt}\mten=\frac{1}{16\pi^2}[4Y_D^2(A_D^2+\dbfive+\mten+\mfive)+6Y_U^2(A_U^2+\dfive+2\mten)-\frac{144}{5}g_5^2M_5^2-2g_{1_A}^2M_{1_A}^2+g_{1_A}^2m^2_{Y}],\nonumber\\
&\frac{d}{dt}\dfive=\frac{1}{16\pi^2}[6Y_U^2(A_U^2+\dfive+2\mten)+30\lamh^2(\ah^2+\dfive+\dfifty+\msi)-\frac{96}{5}g_5^2M_5^2-8g_{1_A}^2M_{1_A}^2-2g_{1_A}^2m^2_{Y}],\nonumber\\
&\frac{d}{dt}\dbfive=\frac{1}{16\pi^2}[8Y_D^2(A_D^2+\dbfive+\mten+\mfive)+30\lamhb^2(\ahb^2+\dbfive+\dfiftyb+\msi)-\frac{96}{5}g_5^2M_5^2-2g_{1_A}^2(4M_{1_A}^2+m^2_{Y})],\nonumber\\
&\frac{d}{dt}\dfifty=\frac{1}{16\pi^2}[3\lamh^2(\ah^2+\dfive+\dfifty+\msi)+2\lams^2(\as^2+\dfifty+\dfiftyb+m_S^2)-\frac{336}{5}g_5^2M_5^2-2g_{1_A}^2(4M_{1_A}^2-m^2_{Y})],\nonumber\\
&\frac{d}{dt}\dfiftyb=\frac{1}{16\pi^2}[3\lamhb^2(\ahb^2+\dbfive+\dfiftyb+\msi)+2\lams^2(\as^2+\dfifty+\dfiftyb+m_S^2)-\frac{336}{5}g_5^2M_5^2-2g_{1_A}^2(4M_{1_A}^2-m^2_{Y})],\nonumber\\
&\frac{d}{dt}\msi=\frac{1}{16\pi^2}[2\lamh^2(\ah^2+\dfive+\dfifty+\msi)+2\lamhb^2(\ahb^2+\dbfive+\dfiftyb+\msi)+\lam_\si^2(A_\si^2+3m_\si^2)-64g_5^2M_5^2],\nonumber\\
&\frac{d}{dt}m_S^2=\frac{1}{16\pi^2}[100\lams^2(\as^2+\dfifty+\dfiftyb+m_S^2)-32g_{1_A}^2M_{1_A}^2+4g_{1_A}^2m^2_{Y}],\nonumber\\
&\frac{d}{dt}Y_U=\frac{1}{16\pi^2}Y_U[9Y_U^2+4Y_D^2+15\lamh^2-\frac{96}{5}g_5^2-3g_{1_A}^2],\nonumber\\
&\frac{d}{dt}Y_D=\frac{1}{16\pi^2}Y_D[3Y_U^2+10Y_D^2+15\lamhb^2-\frac{84}{5}g_5^2-3g_{1_A}^2],\nonumber\\
&\frac{d}{dt}\lamh=\frac{1}{16\pi^2}\lamh[3Y_U^2+\frac{35}{4}\lamh^2+\lamhb^2+\frac{1}{2}\lamsig^2+\lams^2-\frac{188}{5}g_5^2-4g_{1_A}^2],\nonumber\\
&\frac{d}{dt}\lamhb=\frac{1}{16\pi^2}\lamhb[4Y_D^2+\frac{35}{4}\lamhb^2+\lamh^2+\frac{1}{2}\lamsig^2+\lams^2-\frac{188}{5}g_5^2-4g_{1_A}^2],\nonumber\\
&\frac{d}{dt}\lamsig=\frac{1}{16\pi^2}\lamsig[3\lamhb^2+3\lamh^2+\frac{3}{2}\lamsig^2-48g_5^2],\nonumber\\
&\frac{d}{dt}\lams=\frac{1}{16\pi^2}\lams[\frac{3}{2}\lamhb^2+\frac{3}{2}\lamh^2+52\lams^2-\frac{168}{5}g_5^2-12g_{1_A}^2],\nonumber\\
&\frac{d}{dt}A_U=\frac{1}{8\pi^2}[9Y_U^2A_U+4Y_D^2A_D+15\lamh^2\ah+\frac{96}{5}g_5^2M_5+3g_{1_A}^2M_{1_A}],\nonumber\\
&\frac{d}{dt}A_D=\frac{1}{2\pi^2}[3Y_U^2A_U+10Y_D^2A_D+15\lamhb^2\ahb+\frac{84}{5}g_5^2M_5+3g_{1_A}^2M_{1_A}],\nonumber
\end{flalign}
\begin{flalign}
&\frac{d}{dt}\ah=\frac{1}{8\pi^2}[3Y_U^2A_U+\frac{35}{4}\lamh^2\ah+\lamhb^2\ahb+\frac{1}{2}\lamsig^2\asig+\lams^2\as+\frac{188}{5}g_5^2M_5+4g_{1_A}^2M_{1_A}],\nonumber\\
&\frac{d}{dt}\ahb=\frac{1}{8\pi^2}[4Y_D^2A_D+\frac{35}{4}\lamhb^2\ahb+\lamh^2\ah+\frac{1}{2}\lamsig^2\asig+\lams^2\as+\frac{188}{5}g_5^2M_5+4g_{1_A}^2M_{1_A}],\nonumber\\
&\frac{d}{dt}\asig=\frac{1}{8\pi^2}[3\lamhb^2\ahb+3\lamh^2\ah+\frac{3}{2}\lamsig^2\asig+48g_5^2M_5],\nonumber\\
&\frac{d}{dt}\as=\frac{1}{8\pi^2}[\frac{3}{2}\lamhb^2\ahb+\frac{3}{2}\lamh^2\ah+52\lams^2\as+\frac{168}{5}g_5^2M_5+12g_{1_A}^2M_{1_A}],\nonumber\\
&\frac{d}{dt}\mu_\si=\frac{1}{16\pi^2}\mu_\si [2\lamh^2+2\lamhb^2+\lamsig^2-32g_5^2],\nonumber\\
&\frac{d}{dt}B_\si=\frac{1}{8\pi^2}[2\lamh^2\ah+2\lamhb^2\ahb+\lamsig^2\asig+32g_5^2M_5],\nonumber\\
&m^2_{Y}=-5\dfive-5\dbfive+50\dfifty+50\dfiftyb-2m_S^2+\frac{5}{2}\mfive+5\mten.
\label{rgesu5}
\end{flalign}
\bigskip

\bibliographystyle{JHEP}

\bibliography{myref}

\end{document}